\documentclass[usenatbib]{mnras}
\usepackage[varg]{newtxmath}
\usepackage[T1]{fontenc}
\usepackage{ae,aecompl}
\usepackage{graphicx}
\usepackage[figuresleft]{rotating}
\usepackage{lscape}
\usepackage{journals}
\usepackage{comment}
\usepackage{multirow}
\usepackage{rotating}
\usepackage{xcolor,listings}
\usepackage{textcomp}
\usepackage{etoolbox}
\usepackage{hhline}
\usepackage{array}

\definecolor{grey}{rgb}{0.4,0.6,0.6}
\definecolor{brown}{rgb}{0.65,0.16,0.16}
\definecolor{darkgreen}{rgb}{0.0,0.6,0.0}

\defcitealias{ZM11}{ZM2011}

\begin{document}

\title[Galaxy luminosity functions in CGs]{The influence of Hickson-like compact group environment on galaxy luminosities}
\author[Zandivarez, D\'iaz-Gim\'enez \& Taverna]
{A. Zandivarez\thanks{ariel.zandivarez@unc.edu.ar},
E. D\'iaz-Gim\'enez,
A. Taverna
\\
\\
Universidad Nacional de C\'ordoba (UNC). Observatorio Astron\'omico de C\'ordoba (OAC), Laprida 854, X5000BGR, C\'ordoba, Argentina\\
CONICET. Instituto de Astronom\'ia Te\'orica y Experimental (IATE), Laprida 854, X5000BGR, C\'ordoba, Argentina\\
}

\date{\today}
\pagerange{\pageref{firstpage}--\pageref{lastpage}}
\maketitle
\label{firstpage}

\begin{abstract}
Compact groups of galaxies are devised as extreme environments where interactions may drive galaxy evolution.
In this work, we analysed whether the luminosities of galaxies inhabiting compact groups differ from those of galaxies in loose galaxy groups. 
We computed the luminosity functions of galaxy populations inhabiting a new sample of 1412 Hickson-like compact groups of galaxies identified in the Sloan Digital Sky Survey Data Release 16.  We observed a characteristic absolute magnitude for galaxies in compact groups brighter than that observed in the field or loose galaxy systems. We also observed a deficiency of faint galaxies in compact groups in comparison with loose systems. Our analysis showed that the brightening is mainly due to galaxies inhabiting the more massive compact groups. In contrast to what is observed in loose systems where only the luminosities of Red (and Early) galaxies show a dependency with group mass, luminosities of Red and Blue (also Early and Late) galaxies in compact groups are affected similarly as a function of group virial mass. When using Hubble types, we observed that Elliptical galaxies in compact groups are the brightest galaxy population, and groups dominated by an Elliptical galaxy also display the brightest luminosities in comparison with those dominated by Spiral galaxies. Moreover, we show that the general luminosity trends can be reproduced using a mock catalogue obtained from a semi-analytical model of galaxy formation. These results suggest that the inner extreme environment in compact groups prompts a different evolutionary history for their galaxies.
 \end{abstract}

\section{Introduction}
When thinking about extreme environments, compact groups of galaxies (CGs) arise as the perfect candidates. A few bright galaxies close to each other inhabiting a small region of space and relatively isolated from other bright galaxies make them an ideal scenario to study galaxy evolution driven by interactions. The pioneer works of \cite{Hickson82} and \cite{Hickson92} set the basis for identifying compact groups in galaxy surveys. The classic definition establishes four criteria: (1) CGs have between four to ten bright galaxies in a range of three magnitudes from the brightest one (population or magnitude concordance criterion); (2) there is no other bright galaxy in the near surroundings, i.e., within three times the size of the minimum circle that encloses the galaxy members (isolation criterion); (3) the bright member galaxies are within a small projected region making the mean surface brightness of the group to be relatively high (compactness criterion); and (4) the members are within 1000 km/s from the group centre (velocity filter).

Among the different studies that can be done to understand galaxy formation and evolution, the study of the variation of galaxy luminosities in different environments can provide important clues.   
The most commonly implemented procedure to achieve this goal in extragalactic astronomy is the study of the galaxy luminosities through the parameterization of the galaxy luminosity function (LF) proposed by \cite{schechter} using two main parameters: the characteristic absolute magnitude ($M^{\ast}$) and the faint-end slope ($\alpha$).
One of the first determination of the LF of galaxies in CGs for a considerable large sample was performed by \cite{MendesdeOliveira&Hickson91}. They estimated the LF in the B-band using 68 systems extracted from the original sample of Hickson Compact Groups (hereafter HCGs). 
These determinations were performed only for those galaxy members that follow the magnitude concordance selection criterion\footnote{They performed Monte Carlo simulations taking into account the sample selection criteria to estimate the best fit models for the LF.}. The authors found that their results differs from those observed for galaxies in the field or larger galaxy systems. They observed a very strong deficiency of intrinsically faint galaxies ($\alpha \sim 0.2$), as well as elliptical galaxies in CGs being brighter than the average luminosity observed in galaxy clusters. 

Later, using the same sample of 68 HCGs, \cite{sulentic94} estimated the galaxy LF but using a different approach to avoid possible biases due to the Hickson criteria (mainly, magnitude concordance and isolation). They built control samples of galaxy pairs and isolated galaxies that closely mimic the Hickson criteria. Their results showed that CGs have many elliptical galaxies that are brighter than the luminosities observed for isolated galaxies, while only having a few spiral galaxies with no signs of enhanced star formation as seen in galaxy pairs.  In the same year, \cite{ribeiro94} performed a new determination of the LF of galaxies in 22 HCGs, but adding fainter galaxies by performing background discounts. They found a faint-end slope of $\alpha =-0.80\pm 0.15$, which suggested to the authors that there is no pronounced deficiency of faint galaxies in CGs. This result is in agreement with that obtained by \cite{zepf97} for a completed set of 17 HCGs (using a redshift survey) spanning an absolute magnitude range from -23 to -15 in the B-band. They found $M^{\ast}$ and $\alpha$ consistent with those observed for galaxies in the field, in disagreement with the results obtained by \cite{MendesdeOliveira&Hickson91}. 

However, the controversy does not end there. \cite{hunsberger98} used a sample of 39 HCGs with photometric information in the R-band and estimated the galaxy LF. They found that their LF was better described by a double Schechter function, with a $M^{\ast} \sim -21.6$ for the bright end, and a $\alpha \sim -1.17$ for the faint end. This behaviour also showed a clear deficit of intermediate luminosity galaxies around -18. They also found that CGs with a elliptical (or SO) galaxy as first-ranked galaxy showed an excess of faint population compared to those with a first-ranked spiral galaxy. Another similar work was performed by \cite{krusch06}, who estimated the galaxy LF in the B-band using 5 HCGs for which the population of faint galaxies was completed using the red sequence. They also found that the LF is better described by a double Schechter function, but they did not observe a deficiency of galaxies of intermediate luminosities. More recently, \cite{yamanoi20} used new images from the Subaru Hyper Suprime-Cam in the g-band to complete 4 HCGs and estimated their galaxy LF in the absolute magnitude range from -16 to -10. Their results showed a dip in the LF around -12 and a very steep faint-end, leading the authors to hypothesise that probably galaxy interactions may be responsible for the lack of galaxies in the dip, which allows to clearly differentiate two galaxy populations in the LF. 

Beyond the results obtained with the HCG sample, using a sample of 69 CGs identified\footnote{The CG finder selects all sets of galaxies with three or more galaxies within a circular area of ${\rm 200 \ kpc \ h^{-1}}$ radius and a $\Delta v < {\rm 1000 \ km \ s^{-1}}$ \citep{Focardi&Kelm02}. 77\% of the sample are triplets.} in the Updated Zwicky Catalogue \citep{uzc}, \cite{Kelm&Focardi04} compared the LF of galaxies in CGs with those that are in the vicinity of the CGs (within a CG-centred ring that extends between 0.2 to 1 ${\rm Mpc \ h^{-1}}$) or that are completely isolated. 
They observed an excess of early-type and a deficiency of late types when compared with isolated galaxies. They also observed that Elliptical galaxies dominate in number in CGs and are typically brighter than their counterpart in their vicinity.
They found that CGs with an Elliptical or {\bf S0} first-ranked galaxy lack of faint galaxies in comparison with their close environment and are very similar to small clusters. On the other hand, those CGs dominated by an Spiral galaxy resemble our Local Group (one giant spiral and several faints, plus a bright neighbour far away) in large scale (inside a projected radius of 1 ${\rm Mpc \ h^{-1}}$).

As a secondary result of their work, \cite{coenda12} provided the first estimate of the galaxy LF in CGs using a sample with several hundred objects. They used the sample of CGs identified by \cite{McConnachie+09} on the Sloan Digital Sky Survey (hereafter SDSS) Data Release 6 \citep{sdssdr6}. \cite{McConnachie+09} followed the Hickson criteria to identify CGs in projection (population, compactness and isolation).  The CG sample used by \cite{coenda12} was extracted from the so-called catalogue ``A'', which comprises 9713 CGs with three or more members identified in projection in a galaxy sample up to an apparent limiting magnitude in the r-band of 18. Not all the members have redshift measurements, nor the isolation is certified for all the CGs in this sample.
The authors used a subsample of 846 systems with at least one member with $0.06\leq z \leq 0.18$ and magnitudes $14.5\leq r \leq 17.77$, for which only $\sim 58\%$ of the galaxies have measured redshifts. The resulting LF in the $0.1r$-band (in the absolute magnitude range from -24 to -18) computed by \cite{coenda12} gives a  $M^{\ast}-5\log(h)=-21.21$ and a  $\alpha=-1.19$. These results are consistent with a characteristic magnitude brighter than the observed in the field or in loose groups, while the faint-end slope is consistent with the observed for galaxies in the field. Nevertheless, a couple of points are worth noting: firstly, the sample of CGs cannot be considered strictly as Hickson-like CGs, since the selection made by \cite{McConnachie+09} does not ensure that all groups meet the magnitude concordance criterion, therefore systems with their brightest galaxy close to the magnitud limiting of the catalogue (at less than 3 magnitudes from the limit) may not meet the isolation nor the population criteria (see \citealt{DiazGimenez&Mamon10}); secondly, more than 40\% of the potential galaxy members and several galaxies added for the calculation of the LF near the catalogue apparent magnitude limit do not have redshift measurements which introduces considerable uncertainty in the CG membership. 

Finally, the latest estimation of the galaxy LF for a large sample of CGs was performed by \cite{zheng21}. The sample of CGs was identified by \cite{zheng20} in a revised version of the SDSS catalogue \citep{nyusdss} with improved data reduction and redshift completeness. Their identification intends to follow the Hickson criteria (population, compactness, isolation and velocity filtering) but without including the magnitude concordance criterion. They argue that the latter is only relevant when working with purely photometric surveys. The resulting sample of CGs used by \cite{zheng21} comprises 6080 systems with three or more members and apparent magnitudes in the range $14\leq r \leq 17.77$. Their LF determinations in the $0.1r$-band, obtained as a tool to carry out their main study, were performed for two samples of CGs, embedded systems ($M^{\ast}=-21.21$, $\alpha=-0.85$) and isolated systems ($M^{\ast}=-20.74$, $\alpha=-0.57$). However, these determination should be considered with caution when compared with previous results obtained using the original HCG sample or any other sample of Hickson-like CGs because the sample of CGs disregarded the magnitude concordance criterion. Hence, it would be desirable to be able to make a robust determination of the LF of galaxies in CGs using a large, homogeneous sample of systems that fully meets the Hickson criteria. 

A decade ago, \citealt{ZM11} (hereafter \citetalias{ZM11}) performed a very detailed study of the LF of galaxies in loose groups identified in the SDSS Data Release 7 \citep{sdssdr7} as a function of the virial mass of the systems. They confirmed previous findings (e.g., \citealt{zandivarez06,robotham10}) that show a brightening of the characteristic magnitude and a steepening of the faint-end slope of the LF as the virial mass increases. When considering galaxies in the inner or in the outer regions of the systems, the found the same faint-end slopes.
Analysing different galaxy populations, they found that only the luminosities of red spheroids show a strong dependence with group mass, while Late-type galaxy luminosities remain almost invariant. An interesting question would be, how these results observed for the luminosity of galaxies inhabiting loose groups would vary for galaxies that coexist in an environment as hostile as that expected in CGs. Therefore, in this work, we estimate the LF of galaxies in CGs identified in the SDSS DR16 and performed an analysis similar to that performed by \citetalias{ZM11} in loose groups. This new large and homogeneous sample of Hickson-like CGs will enable us to reliably quantify the incidence of such an extreme environment on the luminosities of their galaxy members. 
 
The layout of this work is as follows. In Section 2 we present the different samples of galaxy systems. In Section 3 we compute the LF of galaxies in CGs and compare them with those obtained for loose groups. Finally, in Section 4 we summarise our results and present our conclusions. 

\section{The group samples}
\label{sec:data}
In this section we describe the procedure to identify a new compact group sample using galaxies in a revised sample of the SDSS, as well as a sample of normal or loose groups. 

\subsection{The galaxy sample}
We use the galaxy redshift survey extracted from the SDSS Data Release 16 \citep{dr16}. We select only those galaxies in the main contiguous area of the Legacy Survey. Due to redshift incompleteness of the parent galaxy sample, we complete the selected sample of galaxies using the compilation of \cite{tempel17}\footnote{\url{http://cosmodb.to.ee}} made for the SDSS Data Release 12 \citep{DR12a,DR12b}\footnote{The different catalogues used to complemented this sample are the Two-degree Field Galaxy Redshift Survey \citep{2df1,2df2}, the Two Micron All Sky Survey Redshift Survey \citep{2mass1,2mass2,2mass3}, and the Third Reference Catalogue of Bright Galaxies \citep{RC3a,RC3b}.}. We also included some corrections to the galaxy sample following \cite{DiazGimenez+18},i.e., we removed 21 objects classified as part of a galaxy (PofG) and added 61 new galaxy redshifts obtained from the NASA/IPAC Extragalactic Database (NED)\footnote{\url{https://ned.ipac.caltech.edu/}}.
The extended galaxy sample comprises $565 \, 224$ galaxies with observer-frame model magnitudes $r \leq 17.77$, observer-frame colour $g - r \leq 3$ to avoid stars, and redshifts corrected to the CMB rest frame $z_{\rm CMB} \leq 0.2$ within a solid angle of $6828$ square degrees. All magnitudes are in the AB system and corrected for extinction.   

When needed, k-corrections are computed using the kcorrect\footnote{Version 4.3 extracted from \url{http://kcorrect.org}} code developed by \cite{kcorrect}. In general, we estimate our rest-frame absolute magnitudes using a band shift to redshift 0.1 ($\sim$ the mean redshift of our galaxy sample). We also apply an evolution correction following \cite{blanton2003}. These shifted magnitudes are referred to as $0.1mag$. The cosmological parameters used here are those obtained by the \cite{Planck+14}: $\Omega_m=0.31$ (matter density parameter), $h=0.67$ (dimensionless $z=0$ Hubble constant) and  ${\sigma}_8=0.83$ (standard deviation of the power spectrum on the scale of $8\,h^{-1}\,\rm Mpc$).

\subsection{The compact group sample}
\label{sec:cgs}
To build the CG catalogue from the galaxy sample, we follow a procedure similar to that used by \cite{DiazGimenez+18}.  The complete procedure can be summarised as follows:
\begin{itemize}
    \item We applied the Hickson-like modified algorithm developed by \cite{DiazGimenez+18}. This algorithm takes into account the following criteria: 
    \begin{itemize}
        \item population or magnitude concordance: $4 \le N \le 10$; 
        \item compactness: $\displaystyle \mu_r \le 26.33$; 
        \item isolation: $\displaystyle \Theta_N > 3 \,\Theta_G$; 
        \item flux limit: $\displaystyle r_{\rm b} \le 17.77 - 3$; 
        \item velocity filtering: $ \Delta {\rm v}_{i,{\rm cm}}/(1+z_{\rm cm})  \le  \displaystyle  1000 \, \rm km \, s^{-1} $; 
        \end{itemize}
        where $N$ is the number of members within a three-magnitude range from the brightest galaxy, $\mu_r$ is the group mean surface brightness in the r-band, $\Theta_G$ is the angular diameter of the smallest circumscribed circle that encloses the galaxy members, $\Theta_N$ defines the isolation area where no other bright galaxy within the three-magnitude range is found, $r_b$ is the observer-frame apparent magnitude of the brightest galaxy of the group, $z_{\rm cm}$ is the biweighted median of the group centre, and $\Delta {\rm v}_{i,{\rm cm}}$ is the line-of-sight velocity difference of any galaxy from the biweighted median velocity of the group. 
    This modified version of the algorithm applies the velocity filtering at the same time as the other constraints, i.e., galaxies considered to be neighbours are initially taken from a cylinder in redshift space around the point where the criteria are about to be applied. This procedure improves the completeness of the resulting sample since it avoids discarding groups that might be contaminated by discordant-velocity galaxies when observed only in projection (see subsection 3.3 of \citealt{DiazGimenez+18} for further details). 
    Nevertheless, the implementation of this algorithm in this work to the parent galaxies in the SDSS DR16 differs in two modifications from the former work: firstly, we allowed the inclusion of CGs with a minimum of 3 galaxy members (triplets); and secondly, our algorithm implementation start from the minimum possible number of galaxy members to the maximum allowed (i.e, from 3 to 10). The latter has little impact on the resulting sample, but favours smaller systems even though they may be embedded in a larger configuration that could also meet Hickson's criteria.
    \item The resulting sample of CGs comprises 1662 systems with 5430 galaxy members. 
    \item Given the well known limitations of the SDSS survey regarding fibre collision and the fibre magnitude limit to avoid saturation, as well as objects misclassified as galaxies by the pipeline, we procedeed as follows
    \begin{itemize}
        \item We visually inspected the full sample of galaxies in CGs using the SDSS DR16 Image List Tool\footnote{\url{http://skyserver.sdss.org/dr16/en/tools/chart/listinfo.aspx}}. The inspection showed that $84$ CG members were objects misclassified as galaxies, being actually part of a galaxy (PofG). We list the SDSS \texttt{objID} of these objects in Table~\ref{tab:pofg}. 
        \item We used a photometric sample of SDSS DR16 galaxies to search for galaxies that might be located in the surroundings of each identified CG and were not detected in the spectroscopic survey. Our search only contemplates those galaxies that lie within the isolation disk around each CG, and whose $r$ model magnitudes are within a three-magnitude range from the brightest CG member galaxy. We found $2609$ objects without redshift information that might contaminate our sample of CGs. We visually inspected this list of objects, and searched for alternative spectroscopic determinations using the NASA/IPAC Extragalactic Database (NED). From this search, we found redshift determinations for $159$ galaxies, while $2129$ of those objects did not contaminate the sample, being, for instance, misclassified as galaxies (PofG, stars, etc.). In Table~\ref{tab:newz}, we quoted the redshifts found for the 159 photometric galaxies.
     \end{itemize}
    \item With this new information, we proceeded to run a second identification of CGs. The new sample of CGs comprises 1582 systems. We performed a second cross search with the photometric sample of galaxies looking for potential sources of contamination. In this opportunity, we followed the procedure described by \cite{DiazGimenez+18} to discard galaxies whose photometric redshifts are discrepant with the median spectroscopic redshift of the CG\footnote{According to \citealt{Beck+16}, galaxies with $\displaystyle |z_{\rm phot} - z_{\rm cm} |/(1+z_{\rm cm}) > 0.06$ can be safely discarded as outliers}. We found a list of $253$ galaxies which could make the isolation criterion fail. Finally, we used the photometric properties of those objects and performed a probabilistic analysis in the surface brightness and observer frame $g-r$ colour plane to discard galaxies as potential sources of contamination (see Appendix C of \citealt{DiazGimenez+18} for details of this procedure). As a result, we discard photometric galaxies that lie in the isolation disk but are not really breaking the isolation for 43 CGs. The remaining sample of photometric galaxies, that are within the isolation area of 170 CGs, cannot be ruled out as potential sources of contamination.     
\end{itemize}
Therefore, the final sample of CGs in the SDSS DR16 comprises 1582 systems of which 1412 ($\sim 89\%$) are considered free from any potential source of contamination, while the remaining 170 systems need further spectroscopic information to ensure their isolation. In Appendix~\ref{app:catalogue}, we present the new catalogue of CGs. The list of CGs is presented with a flag that marks the systems with potential sources of contamination. 

Throughout this work we use the sample of 1412 CGs (with 4633 galaxy members) that we consider to satisfy the isolation criterion.

\subsection{The loose group sample}
\label{sec:grp}
We constructed a sample of loose groups from the same parent galaxy catalogue to perform a fair comparison of our results obtained for CGs. 

The algorithm to identify loose systems in a flux-limited catalogue in redshift space is similar to that of \cite{huchra82}.  
This type of algorithm defines galaxy systems by searching for galaxy pairs with projected separations smaller than $D_0 \ R$ and radial velocity differences smaller than $V_0 \ R$, where $R$ is a factor to compensate for the declining galaxy number density as a function of the group distance due to the flux limited survey\footnote{The $R$ factor is $\left(\int_{-\infty}^{M_{12}} \Phi(M) dM/\int_{-\infty}^{M_{lim}} \Phi(M) dM\right)^{-1/3}$ where $\Phi(M)$ is the luminosity function of galaxies in the field, $M_{12}$ and $M_{lim}$ are computed using the mean velocity of the pair and the fiducial velocity ($6000 \ {\rm km \ s^{-1}}$), respectively (both magnitude estimates use the apparent magnitude limit of the catalogue).}. 

In this work, the transverse linking length, $D_0$, is computed following \cite{huchra82} as $D_0= b_{\perp} \ n^{-1/3}$ where $n$ is the mean number density of galaxies\footnote{The mean number density $n$ is computed as $\int_{-\infty}^{M_{lim}} \Phi(M) dM$. The Schechter parameters for the galaxy luminosity function are $M^{\ast}_{0.1r}-5\log(h)=-20.87$, $\alpha=-1.12$ and $\phi^{\ast}=0.01 \ {\rm Mpc^{-3} \ h^3}$. These were estimated for the whole galaxy sample following the procedure described in Sect.~\ref{sec:lfmethod} .} in the Universe and $b_{\perp}=\left(\frac{4\pi}{3}(\frac{\delta\rho}{\rho}+1)\right)^{-1/3}$ is the plane-of-sky dimensionless separation. The later is computed as a function of the number overdensity relative to the mean number density, $\frac{\delta\rho}{\rho}$. Given the adopted Planck cosmology for this work ($\Omega_m=0.315$), an overdensity of $\sim 325$ should be expected \citep{weinberg+03}. Hence, following Planck cosmology we obtain a $b_{\perp}=0.09$, which implies $D_0=0.238 \ {\rm Mpc \ h^{-1}}$ for our loose group identification.  

On the other hand, the radial linking length is estimated using $V_0=b_{\parallel} \ n^{-1/3} \ H_0$ where $b_{\parallel}$ is the line-of-sight dimensionless separation and $H_0=100\ h \ {\rm km s^{-1}/Mpc}$. Following \cite{duarte14}, this equation can be rewritten as $V_0= r \ D_0 \ H_0$ where $r=b_{\parallel}/b_{\perp}= \eta \ \kappa \ \sqrt{\frac{\delta\rho}{\rho} \ \frac{\Omega_{m}}{2}}$ is the analytical prescription for the ratio of line-of-sight to plane-of-sky group sizes due to redshift distortions. Assuming that the velocities in the line-of-sight span the range $\pm \kappa \sigma_v$ with $\sigma_v$ the velocity dispersion, we can set $\kappa=2.58$ to ensure that 99\% of galaxy velocities are included under a one dimensional Maxwellian velocity distribution. Hence, using $\eta \sim 0.65$ \citep{Mamon+13}, we obtain $r \sim 12$, which implies $b_{\parallel}=1.1$ and $V_0=285 \ {\rm km \ s^{-1}}$. 

The final sample of loose groups comprises 14652 objects with four or more members and virial masses greater than $10^{12} \ {\rm {\cal M}_{\odot} \ h^{-1}}$. 

\section{The luminosity function of galaxies in CGs}
\label{sec:lf}
In this section we use the sample of CGs to estimate the LF of their galaxy members. Our estimations will not be limited only to the complete sample of galaxies in CGs, we will investigate the different populations of galaxies characterised by colours, light concentrations and morphologies. We will also analyse the dependence of the results on the mass estimations of our CGs.

\subsection{The LF estimators}
\label{sec:lfmethod}
Two different methods are used to estimate and characterise the LF of galaxies in CGs. The estimation of the binned LF is obtained using the non-parametric $C^{-}$ method developed by \cite{cmethod}. \cite{willmer97} stated that this method is probably the most robust estimator of the LF being less affected by the faint range of luminosities or sample size. On the other hand, we also adopted the method known as STY developed by \cite{STY}, which is a very reliable maximum likelihood method for fitting analytic functions without binning the data. The adopted analytic function for the STY method is the well-known Schechter function \citep{schechter}, which is described by two shape parameters, the characteristic absolute magnitude ($M^{\ast}$) and the faint end slope ($\alpha$). For the search of the maximum in the likelihood surface defined by the STY method, we choose to implement a Markov Chains Monte Carlo (MCMC) method that maps the parameter space using a Metropolis-Hastings algorithm \citep{metropolis,hastings}.  

\subsection{LF of the full sample of galaxies in CGs}

\begin{figure}
\centering
\includegraphics[width=0.236\textwidth]{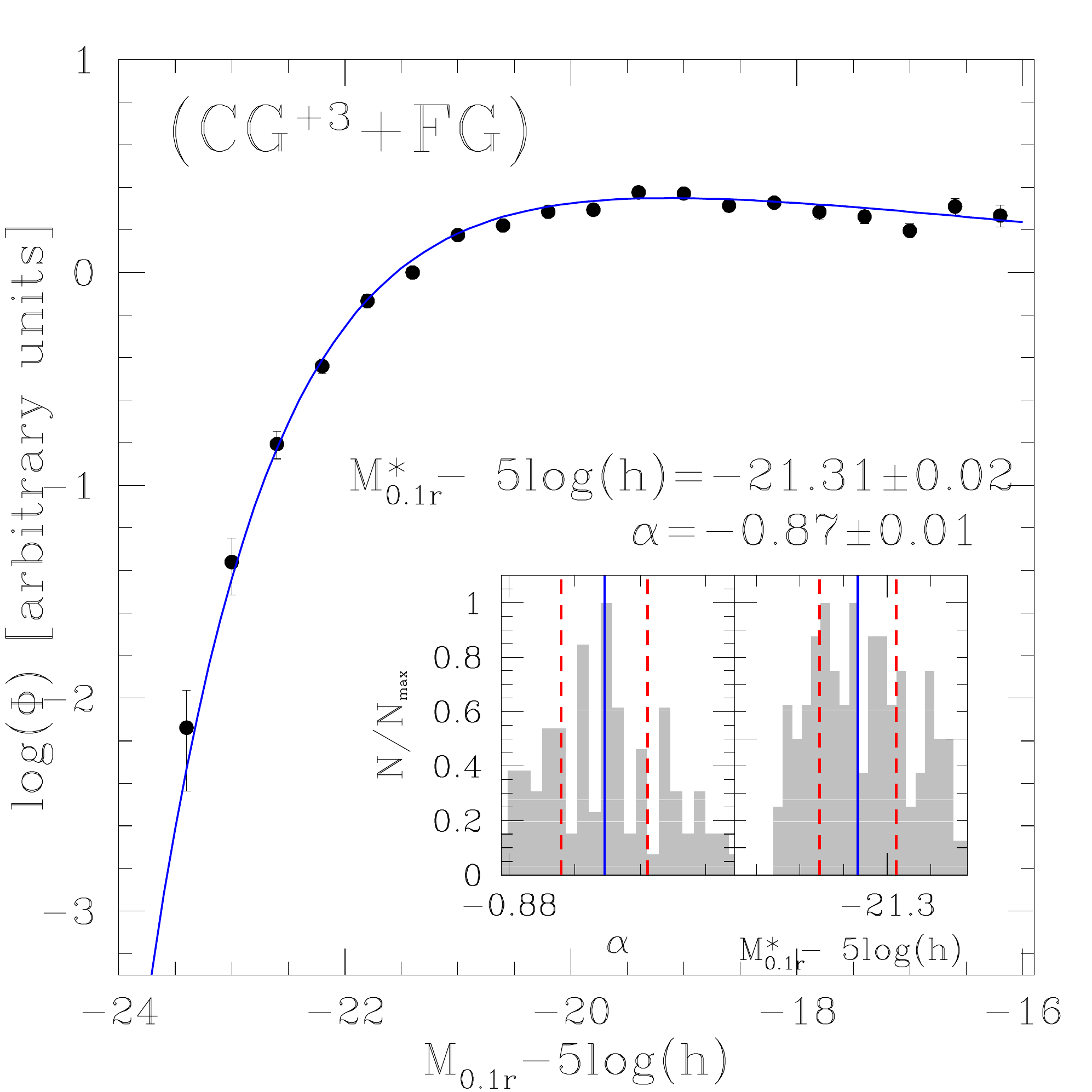}
\includegraphics[width=0.23\textwidth]{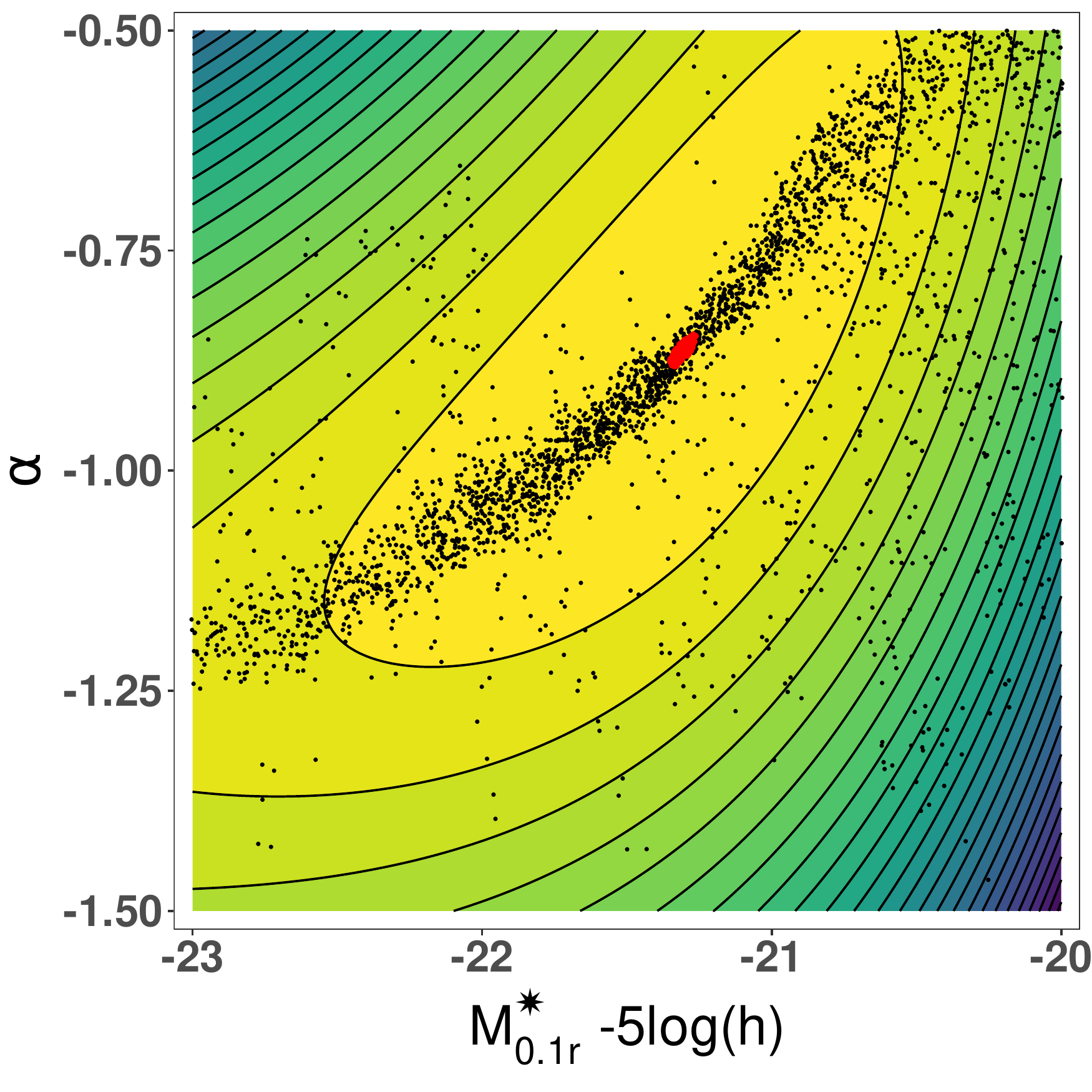}
\includegraphics[width=0.236\textwidth]{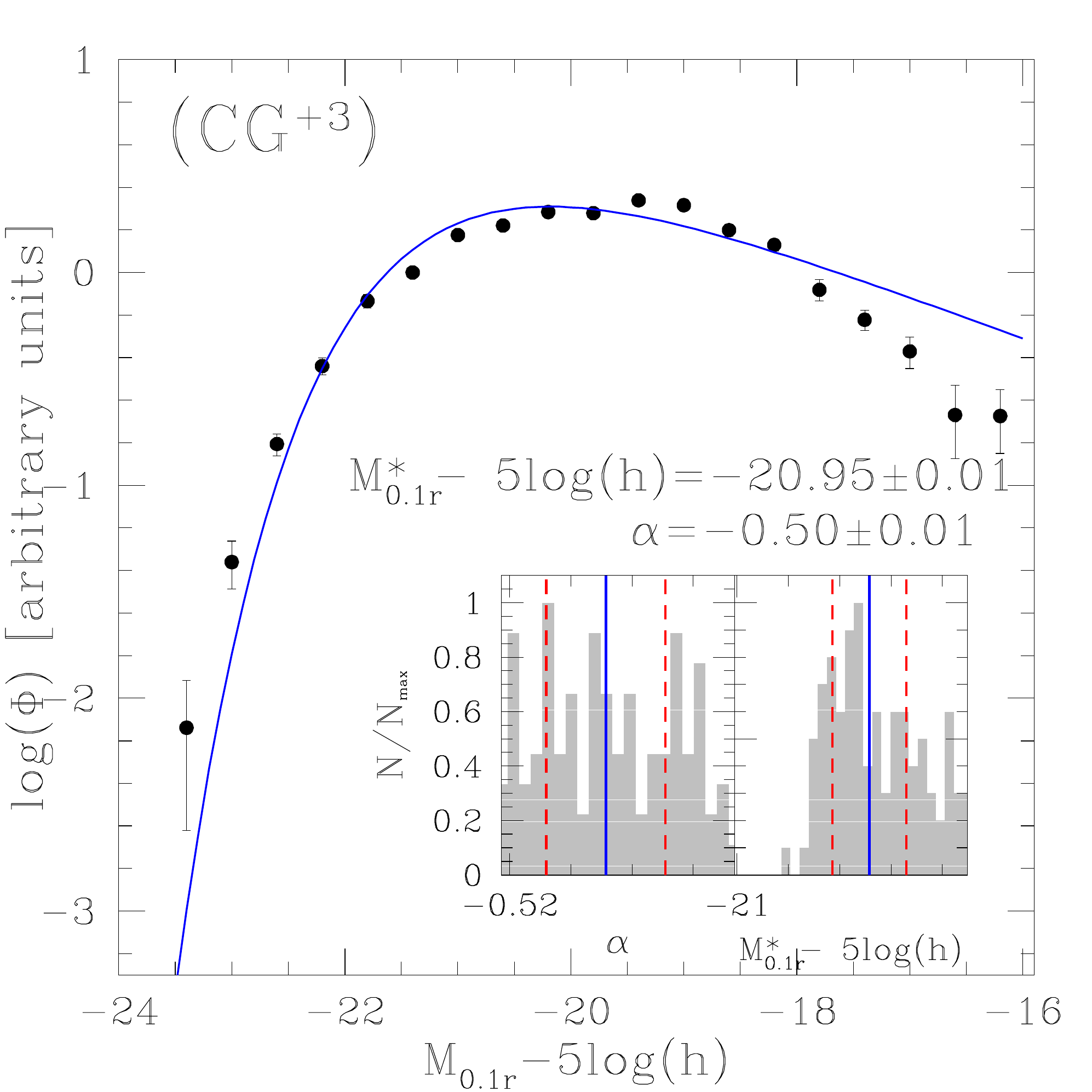}
\includegraphics[width=0.23\textwidth]{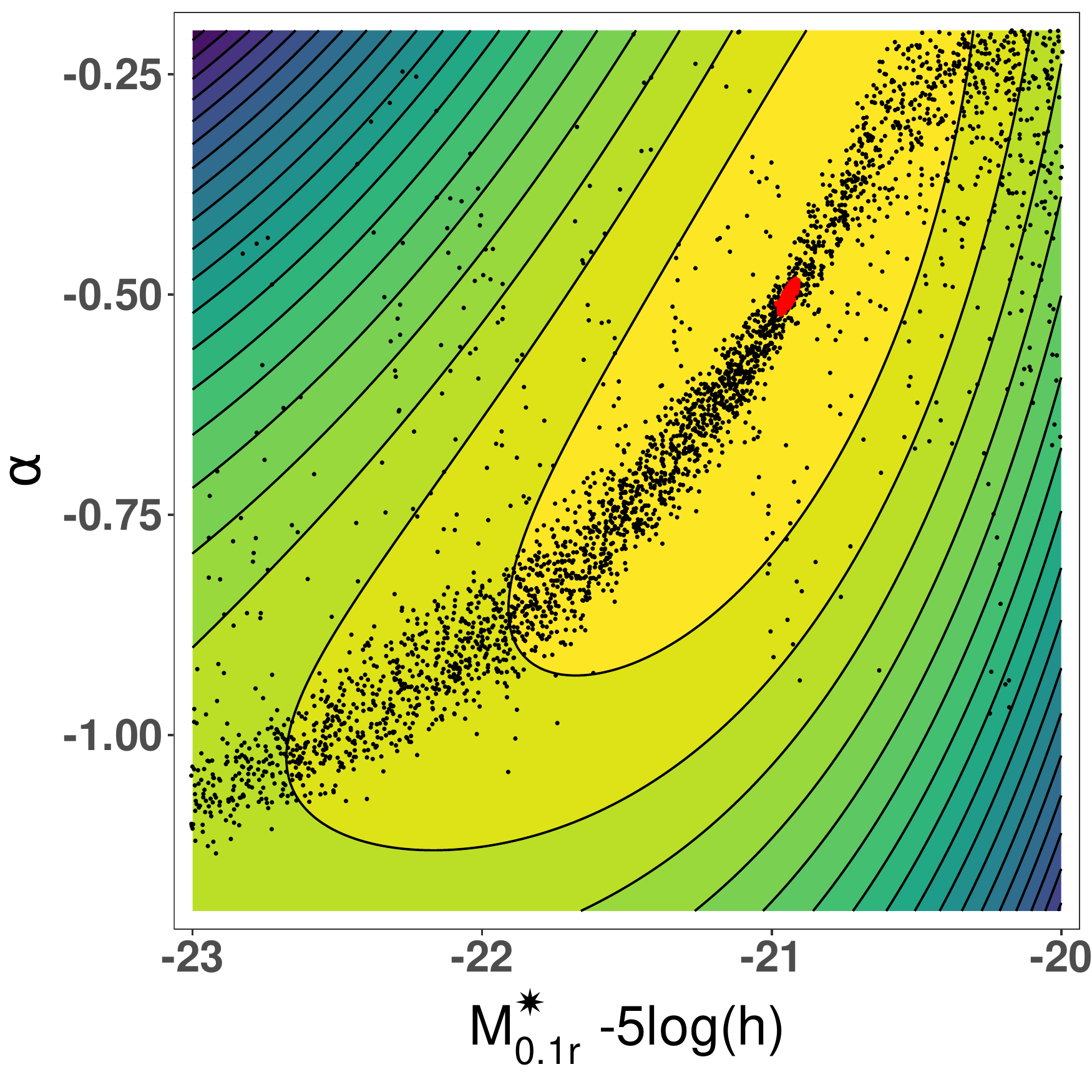}
\includegraphics[width=0.236\textwidth]{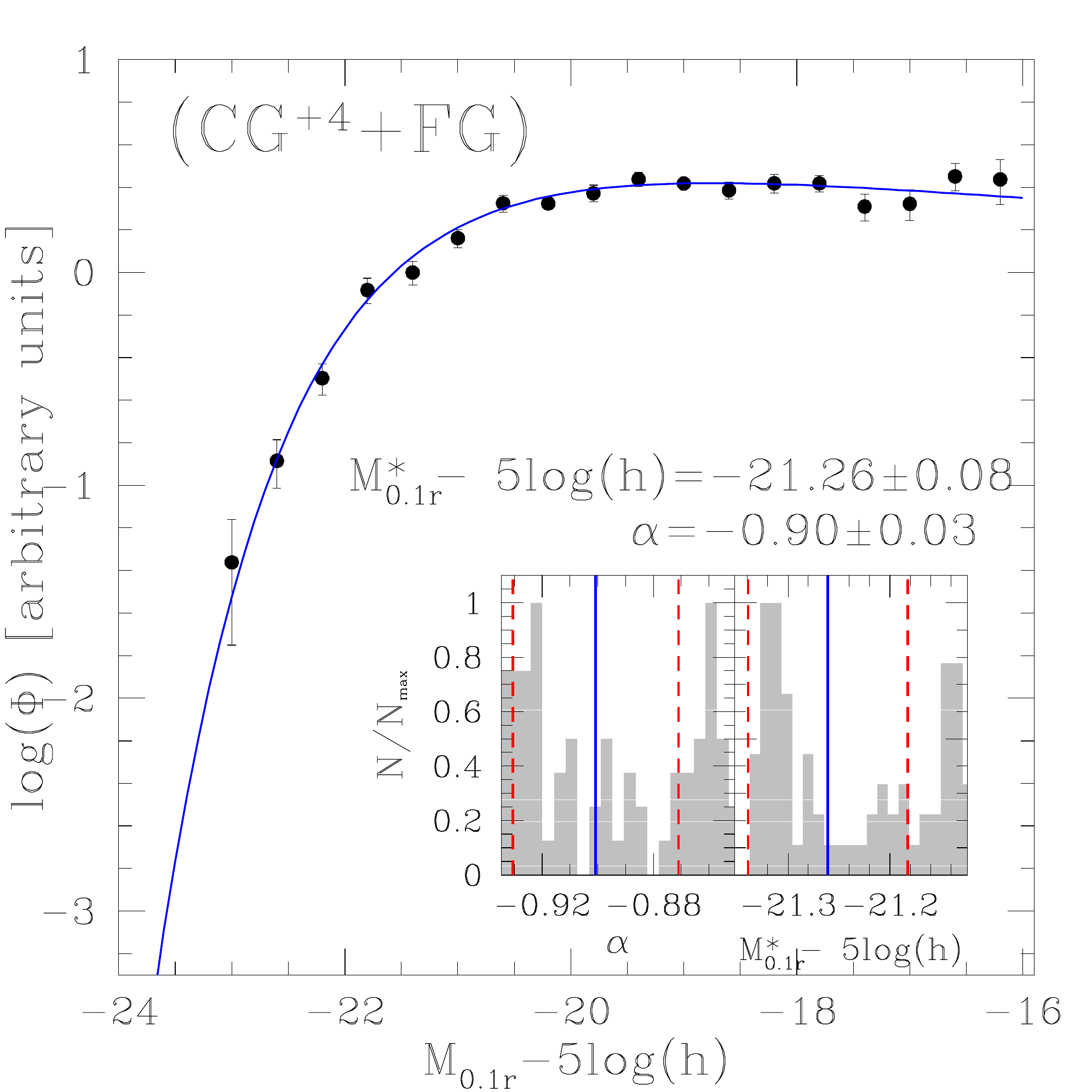}
\includegraphics[width=0.23\textwidth]{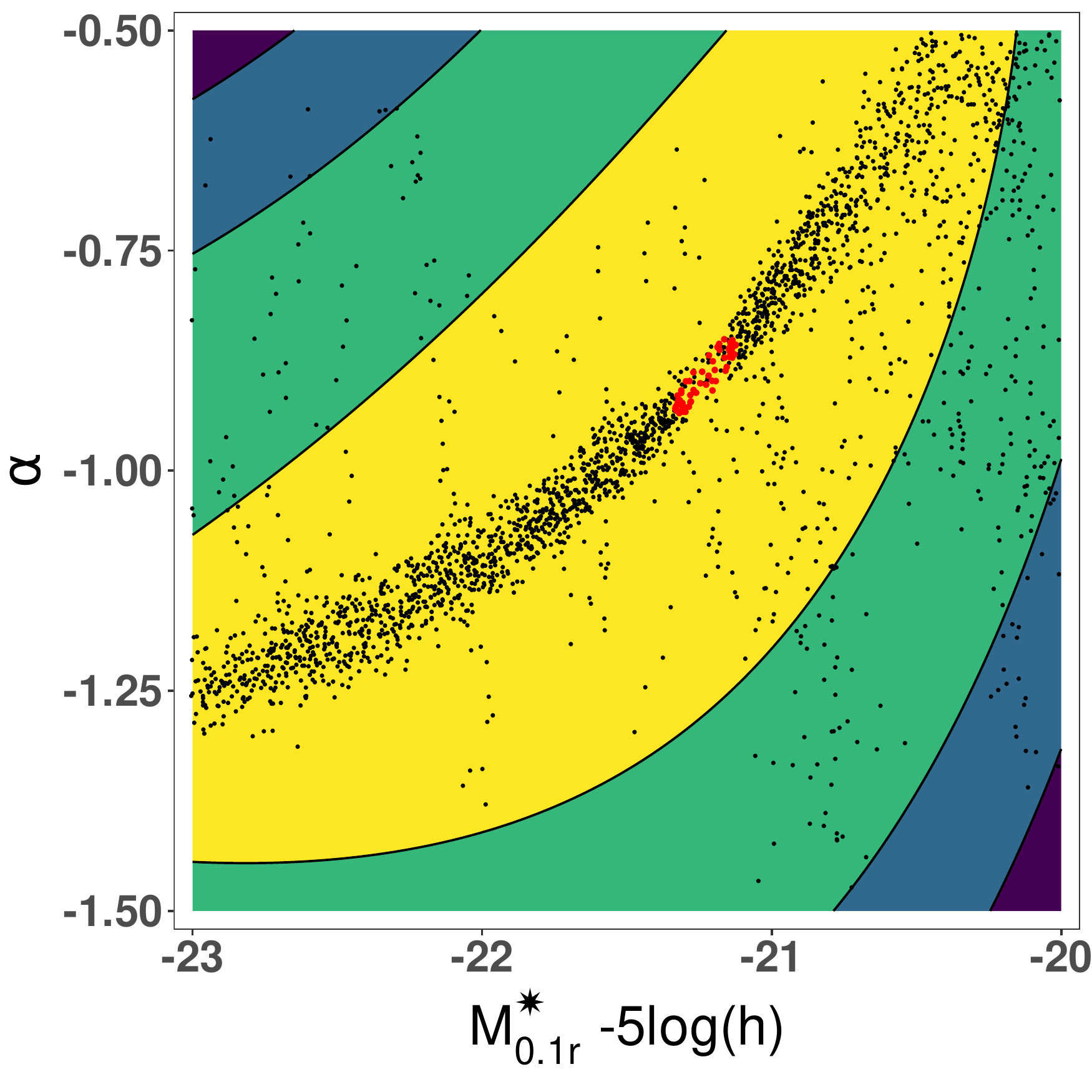}
\caption{Luminosity functions of galaxies in CGs. Left panels show the LFs of galaxies in CGs in the $^{0.1}$r-band. Points show the LFs estimated using the $C^{-}$ method (in arbitrary units), while error bars are estimated from a bootstrap resampling technique. Inset panels show the distribution of STY best-fitting Schechter functions parameters obtained as the end points of 100 markov chains run in the likelihood space. Vertical blue lines in the inset panels show the medians values while the vertical red dotted lines show their semi-interquartile ranges. The corresponding best-fitting Schechter values (vertical blue lines) are quoted in the main panel and the Schechter LF obtained from them is drawn in blue solid line. Right panels show the STY likelihood space obtained for each sample. Black dots represent each point of the 100 markov chains while red dots indicate the end points of those chains. Top panels are constructed using the sample of CGs with three or more bright galaxy members plus the faint galaxies in the CGs isolation volume (see text). In the middle panels the same sample of CGs described above is used but excluding the faint members. Finally, bottom panels are constructed using the sample of CGs with four or more bright galaxy members plus their faint galaxy companions.}
\label{fig:lf_tot}
\end{figure}

An important constraint of CGs is that they are defined as systems where every galaxy member is within a three magnitude range from the brightest galaxy. This is the magnitude concordance criterion defined by \cite{Hickson82}, and it was originally intended to avoid strong distance-dependent biases in a search of systems in projection, but also avoids finding systems formed by just one massive galaxy and their satellites.
The application of this criterion does not mean that there are no faint galaxies in the region where a CG is located, but they have not been used to define it. Disregarding this issue (i.e., using only the CG members obtained by the definition of population) might introduce a luminosity bias, preventing a reliable determination of the true LF of galaxies in CGs.

Therefore, for a proper estimation of the LF, we complete the galaxy sample in CGs by adding the faint galaxies from the parent galaxy catalogue that are located within the CG influence region or isolation cylinder, i.e., those with angular separations from the CG centre inside the isolation disk ($3 \,\Theta_G$) and line-of-sight velocity within $1000 \, \rm km \, s^{-1}$ from the group centre. We found a sample of 1068 faint galaxies to add to the CG galaxy members obtained by definition. Notice that, on average, there are less than 1 or 2 additional faint galaxies per CG. The resulting sample of galaxies in CGs comprises 5701 galaxies. 

Figure \ref{fig:lf_tot} shows the LF for the full sample of galaxies in CGs. Top panels show the estimation of the LF for all the galaxies in CGs with 3 or more members (CG$^{+3}$) plus their corresponding faint galaxies (FG). Top-left panel shows the $C^{-}$ determination (points) and the best-fitted Schechter function (blue line). The corresponding Schechter parameters $\alpha$ and $M^{\ast}$ (in the $0.1r$ band) computed with the STY method are quoted therein. These parameters are the median values of the end points of 100 MCMC in the parameter space (see inset panels, where blue lines are the median values and red dashed lines are the semi-interquartile range used as the parameters errors). Top-right panel illustrates the search for the maximum on the likelihood surface using the 100 MCMC (chains - black dots; end points - red points). 
Middle panels show the LF estimated only with those relatively bright galaxies in CGs defined as members by the searching algorithm (3-magnitude range). Disregarding the faint population of galaxies produces an underestimation of both, the faint-end slope (almost $0.4$) and the resulting characteristic magnitude ($\sim 0.35$). Finally, we computed the LF of CGs (with both, main and faint galaxies) this time excluding triplets from our sample (CG$^{+4}$). This test is presented since several catalogues of CGs do not include triplets. Bottom panels show that the estimation of the LF of galaxies in CG$^{+4}$ is in complete statistical agreement with the determination obtained for the galaxies in CG$^{+3}$ (top panels). This observed consistency encouraged us to carry out the rest of the analysis in this work with the complete sample of CGs, which will allow us to obtain a higher statistical confidence.     

We observe that CGs are characterised by a very bright $M^{\ast}_{0.1r}-5\log(h)=-21.31 \pm 0.02$. Compared with previous determinations for field galaxies in the SDSS, such as \cite{blanton2003} ($M^{\ast}_{0.1r}-5\log(h)=-20.44\pm 0.01$ and $\alpha=-1.05\pm0.01$) or our own estimates for field galaxies in the galaxy sample used in this work ($M^{\ast}_{0.1r}-5\log(h)=-20.87\pm 0.01$ and $\alpha=-1.12\pm0.01$), the characteristic absolute magnitude for galaxies in CGs is between half and one magnitude brighter.
Moreover, the faint-end slope also shows important differences. Our results are better described by a downward slope while an almost flat behaviour was previously observed for field galaxies. 
In addition, our estimate also differs from those previously done in loose galaxy groups. Even though previous works showed that galaxies in loose galaxy groups are brighter than in the field, our findings show that the characteristic magnitude is even brighter for galaxies in CGs. 
For instance, from the work of \citetalias{ZM11} it can be inferred an average $M^{\ast}_{0.1r} - 5\log(h) \sim -21$ and $\alpha \sim -1.15$. 
Therefore, galaxies in CGs seem to be brighter than those inhabiting common galaxy groups, and the differences observed in the faint-end slope are similar to those observed in the field. 
Beyond this rough comparison, a more appropriate and detailed study between CGs and loose groups is performed in the following sections.

We find that the Schechter parametrization is a very good descriptor for the LFs studied below (as an example, see Fig.~\ref{fig:mass_fits} in the Appendix \ref{app:lfparam}), therefore, our findings will be expressed only in terms of the Schechter's shape parameters $M^{\ast}$ and $\alpha$.

\subsection{The group mass dependence of the LF}
\label{sec:lfmass}

We analyse the dependence of the LF on group mass. This allows us to perform a comparison with the results obtained for loose galaxy groups in the SDSS.

\begin{figure}
\centering
\includegraphics[width=0.5\textwidth]{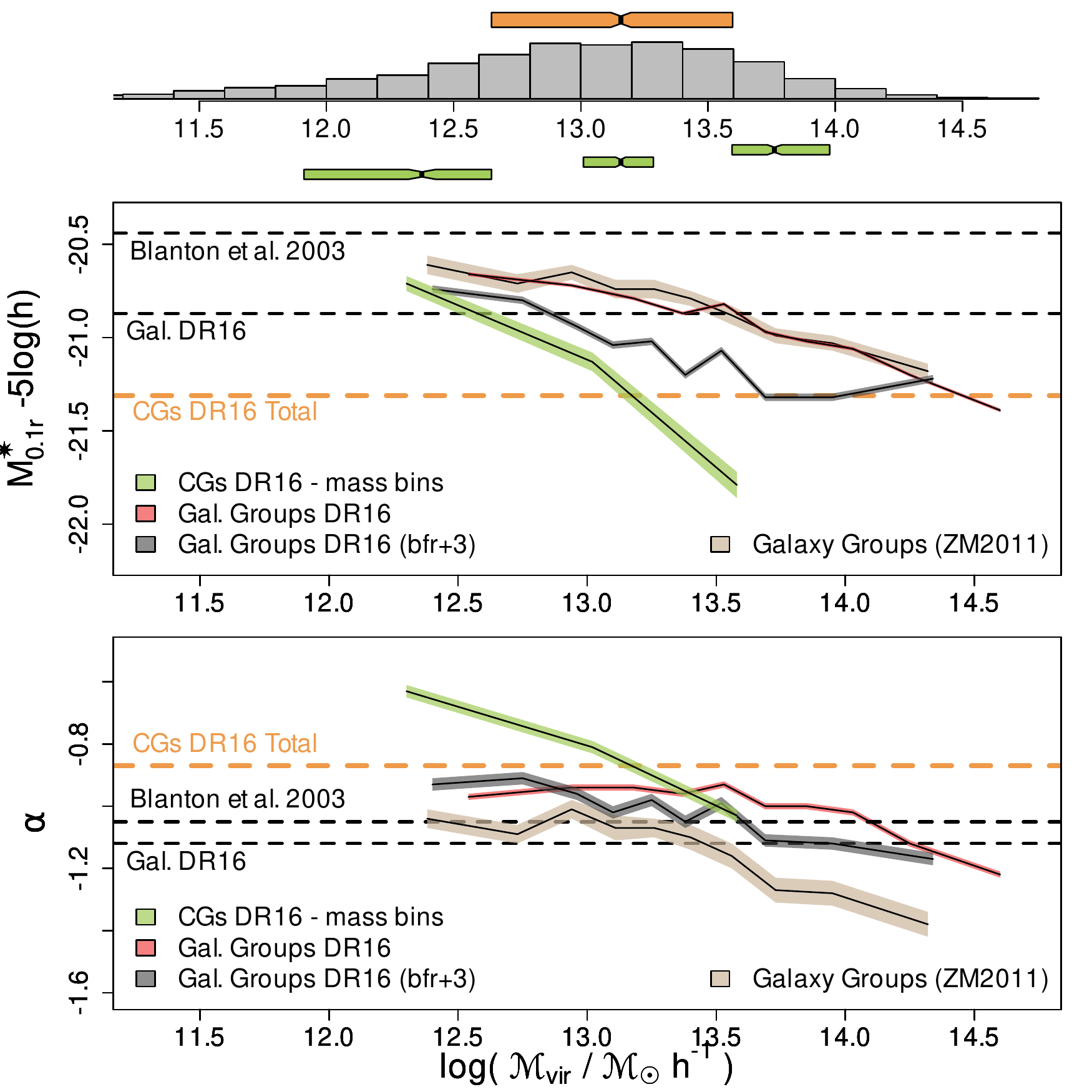}
\caption{The mass dependence of LFs for the total sample of galaxies CGs. Top histogram shows the CGs virial mass distribution: upper orange boxplot shows the CG median virial mass as well as the median 95\% confidence interval (notches) and the inter-quartile range of the distribution; green boxplots show the same quantities but for three equal number CGs subsamples used to estimate the Schechter LF parameters. Middle and bottom panels show the STY best-fitting Schechter LF parameters as a function of CG virial masses. Shaded areas are the semi-interquartile range for the median values of $\alpha$ and $M^{\ast}$ obtained from the markov chains. For comparison, we included the estimations obtained for galaxies in loose galaxy groups identified in the SDSS by \citetalias{ZM11} as well as new estimates for loose groups identified on the DR16 galaxy sample used in this work. The subsample labelled as bfr+3 mimic the flux limit of the brightest and magnitude concordance criteria of CGs (see text in  Sect.~\ref{sec:lfmass}).
Horizontal orange lines are the CGs LF parameters shown in the upper plot of Fig.\ref{fig:lf_tot}, while horizontal black lines are the LF parameters for galaxies in the field obtained for DR16 galaxy sample.}
\label{fig:lf_mass}
\end{figure}

Following \citetalias{ZM11}, group virial masses are computed as ${\cal M}=\sigma^2R_{\rm vir}/G$ \citep{limber60}\footnote{\cite{limber60} used the prescription to estimate the virial mass of the Stephan's Quintet.}, where $R_{\rm vir}$ is the 3D virial radius of the system, and $\sigma$ ($\sqrt{3}\sigma_v$) is the velocity dispersion of member galaxies, while the line-of-sight velocity dispersion $\sigma_v$ is estimated using the biweight ($N\ge 15$) and gapper ($N<15$) estimators described by \cite{Beers+90}. 
We estimate the CG virial masses adopting as galaxy members all galaxies (bright and faint) inhabiting the isolation cylinder of CGs. Histogram at the top of Fig.\ref{fig:lf_mass} shows the distribution of the logarithm of the group virial mass distribution for the full sample of CGs. The left and right edges of the upper orange boxplot represent the 25th and 75th percentiles of the distribution, while the wrist represents the median of the distribution. We can see that the median logarithmic virial mass for the full sample of CGs is $13.02\pm 0.04$ (error is the 95\% confidence interval, shown as notches around the wrist in the boxplot). 

To study the dependence of the LF with the group mass and with the aim of keeping our results statistically reliable, we split the full sample of CGs into three equal number subsamples using the 33th and 66th percentiles of the virial mass distribution (12.73 and 13.28, respectively). The median of the logarithmic values of virial masses for each subsample are $12.30\pm 0.05$, $13.02\pm0.02$ and $13.58\pm0.03$. Using these subsamples of CGs we estimate the LF of galaxies in CGs in the same way as in the previous section. Middle and bottom panels in Fig.\ref{fig:lf_mass} show the STY best-fitting Schechter parameters ($M^{\ast}$ and $\alpha$, respectively) as a function of group virial mass (green lines). A clear dependence of the LF parameters on the CG virial mass is found, with a variation of 1.08 magnitudes in $M^{\ast}$ and a difference of 0.4 in $\alpha$ between the lowest and the highest mass bin.

We also display in Fig.~\ref{fig:lf_mass} the results published by \citetalias{ZM11} (light brown region) for SDSS DR7 loose galaxy groups\footnote{All the Schechter parameters values of this previous work were extracted from Table A1 in \citetalias{ZM11}}. 
However, the flux limit criterion applied on CGs ($r_b\le 14.77$) could introduce a luminosity bias when comparing the results from CGs with those obtained previously for loose groups. In the same sense, the magnitude concordance criterion applied to CGs could also introduce a luminosity bias selection.
Therefore, to perform a fair comparison, we use the sample of loose groups identified on the DR16 galaxy sample in Sect.~\ref{sec:grp}, and restrict it to select only those systems whose brightest galaxy is brighter than 14.77 (meeting the CG flux limit criterion), and also selecting those loose groups with at least three galaxies within a three magnitude range from the brightest galaxy member (mimicking the CG concordance criterion). 
We will refer to this sample as the loose groups with a bright first ranked galaxy and a minimum population within the 3 magnitude range (hereafter, bfr+3), which comprises 2751 loose groups.
The estimates of the parameters of the LF for galaxies in both samples of DR16 loose groups (total and bfr+3) are also shown in Fig.~\ref{fig:lf_mass} (red and grey lines, respectively). 

When comparing the LF parameters of the total sample of DR16 loose groups with those obtained by \citetalias{ZM11} in DR7 (light brown vs. red lines), we observe that the trends for $M^{\ast}$ as a function of mass are quite similar, while the trends obtained for $\alpha$ in DR16 is flatter than the obtained previously in DR7. 
These differences can be due to several factors such as the different procedure applied on each data release (DR7 and DR16) for processing imaging and spectra, the inclusion of new redshift determinations for bright galaxies, or the different group identification processes applied in both works.  

On the other hand, when comparing the characteristic absolute magnitude obtained for the total and the bfr+3 samples of loose groups in DR16 (red vs. grey lines), a brightening of $\sim 0.23$ magnitudes in the restricted sample is observed in the whole range of the group virial masses. Also, a smaller dependence of the faint-end slope is observed as a function of group mass for the bright first ranked loose group sample, keeping its value roughly in the range $-1.0$ and $-1.15$. Hence, there exist differences between the samples with or without the CGs restrictions. Therefore, the comparison with the galaxies in CGs must be done using the restricted sample of loose groups. 

\begin{figure*}
\centering
\includegraphics[trim=0 -2cm 0 0, width=0.40\textwidth]{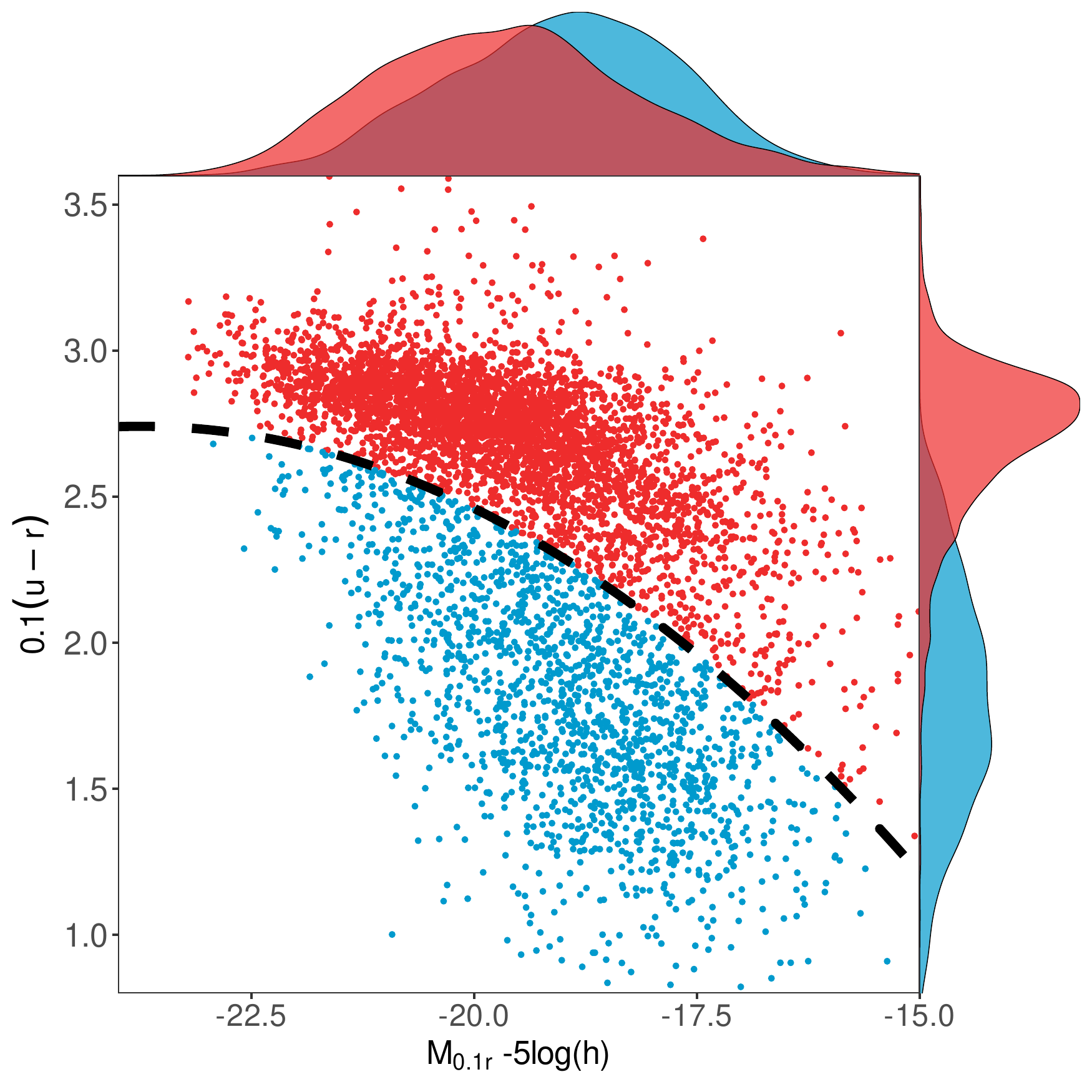}
\includegraphics[width=0.50\textwidth]{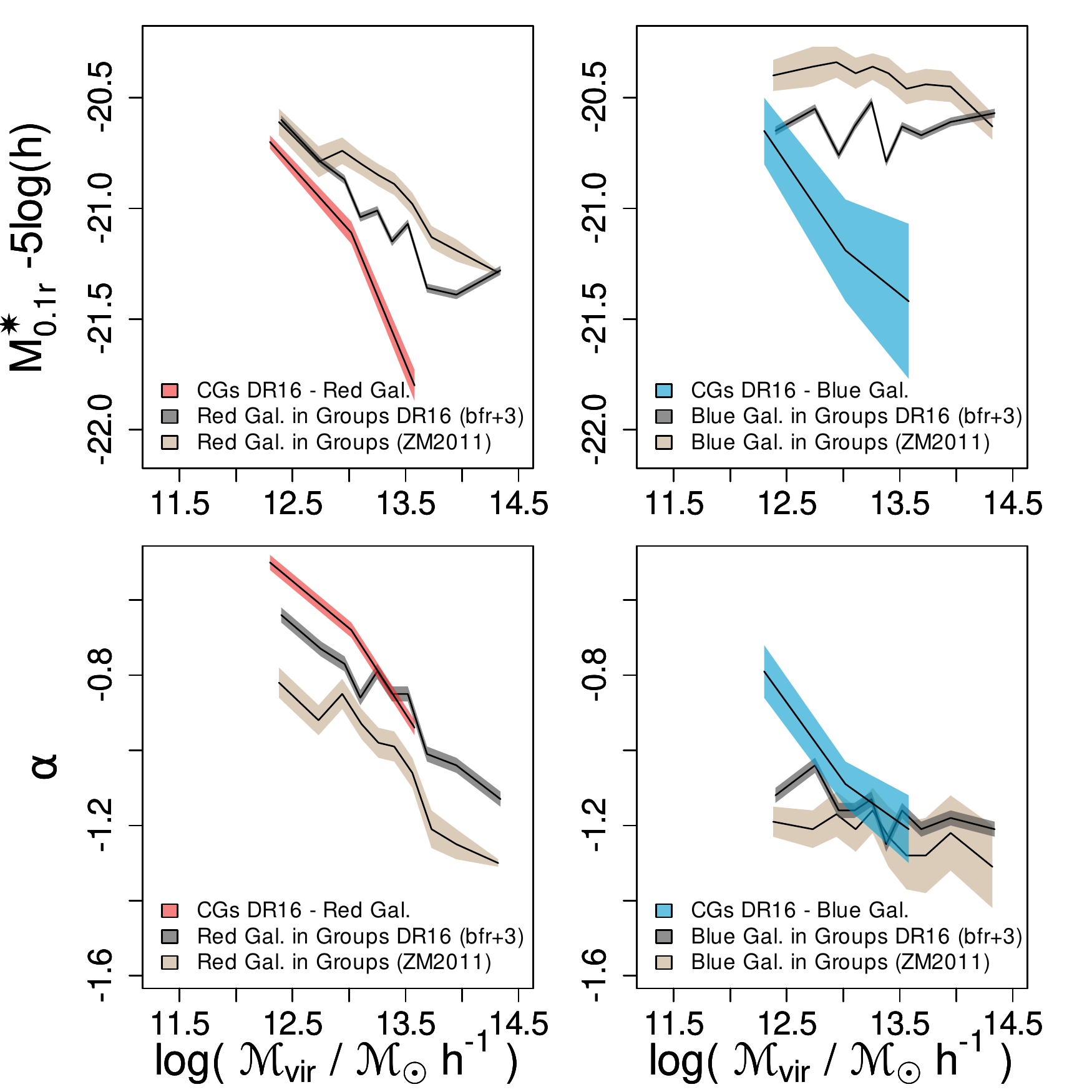}
\caption{The mass dependence of the LFs for the Blue and Red galaxy population in CGs. The left plot shows the colour–magnitude diagram of CG galaxy members. The black dashed line is the function defined by \citetalias{ZM11} to split galaxies into Red (above the curve) and Blue (below the curve) subsamples (see text for details). We also show the marginal distributions at the top and right of the main panel. The four panels at the right show the STY best-fitting Schechter LF parameters obtained for the Red and Blue CG galaxy subsamples as a funtion of CGs virial masses. For comparison, we included the estimations obtained for similar subsamples of Red and Blue galaxy populations in loose groups identified in the SDSS DR7 by \citetalias{ZM11}  and our own estimates for loose groups in the SDSS DR16 with a bright first-ranked galaxy and with at least 3 galaxy members that are at most three magnitudes fainter than the brightest galaxy in the group (bfr+3).}
\label{fig:color_mag}
\end{figure*}

From the comparison between galaxies in CGs and in loose groups (green vs. grey lines in Fig.~\ref{fig:lf_mass}), we observe interesting differences. As a general trend, we observe that galaxies in low mass loose groups show $M^{\ast}$ and $\alpha$ very similar to those corresponding to field galaxies (black dashed lines), while a brightening is observed as group virial mass increases. 
On the other hand, CGs with masses less than $10^{13} \ {\rm {\cal M}_{\odot} \ h^{-1}}$ show a small brightening in $M^{\ast}$ compared to those observed for loose groups, while a most notable brightening is observed in the highest CG mass bin, leading to a difference of almost half a magnitude. 
Also, the faint-end slope of the LFs shows a very different behaviour between galaxies in CGs and in loose groups. A very pronounced downward steepness is observed for low mass CGs that differs in 0.3-0.4 from the obtained for galaxies in loose groups and in the field, while the faint-end slope turns to flat for the highest CG mass bin, similar to the observed for field galaxies and loose galaxy groups. 

These results suggest that there is a very distinctive behaviour in the luminosities of galaxies in CGs. Galaxies inhabiting more massive CGs are typically brighter than in any other galaxy group, and there is a clear deficiency of faint galaxies in low mass CGs compared to the observed in loose galaxy systems.

\subsection{LFs for different galaxy populations}
To deepen the comparison of galaxies in CGs and in loose galaxy groups, we analyse the behaviour of different galaxy populations. 
There is a large number of works in the literature showing that some galaxy properties can correlate with galaxy morphology (e.g. \citealt{humason36,hubble36,morgan57,devauc61,strateva2001,shimasaku2001}). In this work, we choose the $u-r$ colour and the concentration index as indicators of the morphological type. We also complement our analysis using a morphological classification based on Hubble galaxy types obtained from the Galaxy Zoo database. 

\subsubsection{Red and Blue galaxy populations}
The well-known bimodality of galaxy colours is used in this work to split galaxies into red and blue populations. This bimodality depends on galaxy luminosity \citep{baldry04}. Therefore, Red and Blue sequences can be modelled by combining two Gaussian functions to the colour distribution for a given galaxy luminosity. 

We used the rest-frame $0.1(u-r)$ colour\footnote{$0.1(u-r)=(u-k_u^{0.1})-(r-k_r^{0.1})$}, estimated from the model apparent magnitudes\footnote{\cite{baldry04} showed that model magnitudes give higher S/N measurements than the Petrosian magnitudes, turning the model magnitudes the proper choice to estimate galaxy colours.}. Following \citetalias{ZM11}, we  split galaxy populations using their fitted polynomial prescription for colour, $P(x) = -0.02x^2-0.15x+2.46$\footnote{The polynomial fit gives a recipe to obtain, for each luminosity bin, the colour value at which the two Gaussian functions intersect.}, where $x = M_{0.1r} - 5\log(h) + 20$. Galaxies with colours above $P(x)$ are classified as red galaxies, otherwise are blue galaxies.
Left plot of Fig.\ref{fig:color_mag} shows the colour-magnitude diagram for galaxies in CGs. This plot shows the two galaxy populations split using the $P(x)$ prescription of \citetalias{ZM11} (black dashed line). We show the marginal distributions of colours and absolute magnitudes of the Red and Blue galaxy populations at the right and top of the plot, respectively. The density distributions shown in the right show a clear separation in colours among both populations, while a small shift to brighter luminosities is observed in the upper distributions for the Red galaxies.  

Four right panels in Fig.\ref{fig:color_mag} show the Schechter parameters for Red and Blue galaxy populations inhabiting CGs as a function of the group virial masses. We also add the determinations obtained for the same galaxy populations in loose groups (tabulated values for loose groups in DR7: light brown lines, bfr+3 subsample of loose groups in DR16: grey lines). 
Red galaxies inhabiting CGs and in loose galaxy groups behave very similarly as the complete samples of galaxies shown in Fig.\ref{fig:lf_mass}, showing a dependence of the LF parameters as a function of the group masses. Also, red galaxies, mainly in high mass CGs, are notoriously brighter than those in loose galaxy groups, and there is also a lack of faint red galaxies in low mass CGs.
On the other hand, blue galaxies in loose groups show no dependence in the LF parameters with group mass. However, blue galaxies in CGs do. 
While blue galaxies in loose groups are typically less luminous than their red companions, blue galaxies in CGs are as bright as the red galaxies in CGs (in terms of $M^{\ast}$). The faint-end slope of blue galaxies in CGs indicates a lack of blue faint galaxies in low mass CGs, becoming more similar to loose groups in the highest mass bin. 

\subsubsection{Early and Late galaxies}
According to \cite{strateva2001}, the concentration index is an useful proxy for the mass distribution using the stellar light of galaxies, and together with galaxy colour are independent, quantitative indicators of morphology. A similar statement was done by \cite{shimasaku2001}, showing that the (inverse) concentration index is a very suitable tool to classify Early (E/SO) and Late (S/Irr) galaxy types, despite a small contamination compared to visual determinations.

\begin{figure*}
\centering
\includegraphics[trim=0 -2cm 0 0, width=0.40\textwidth]{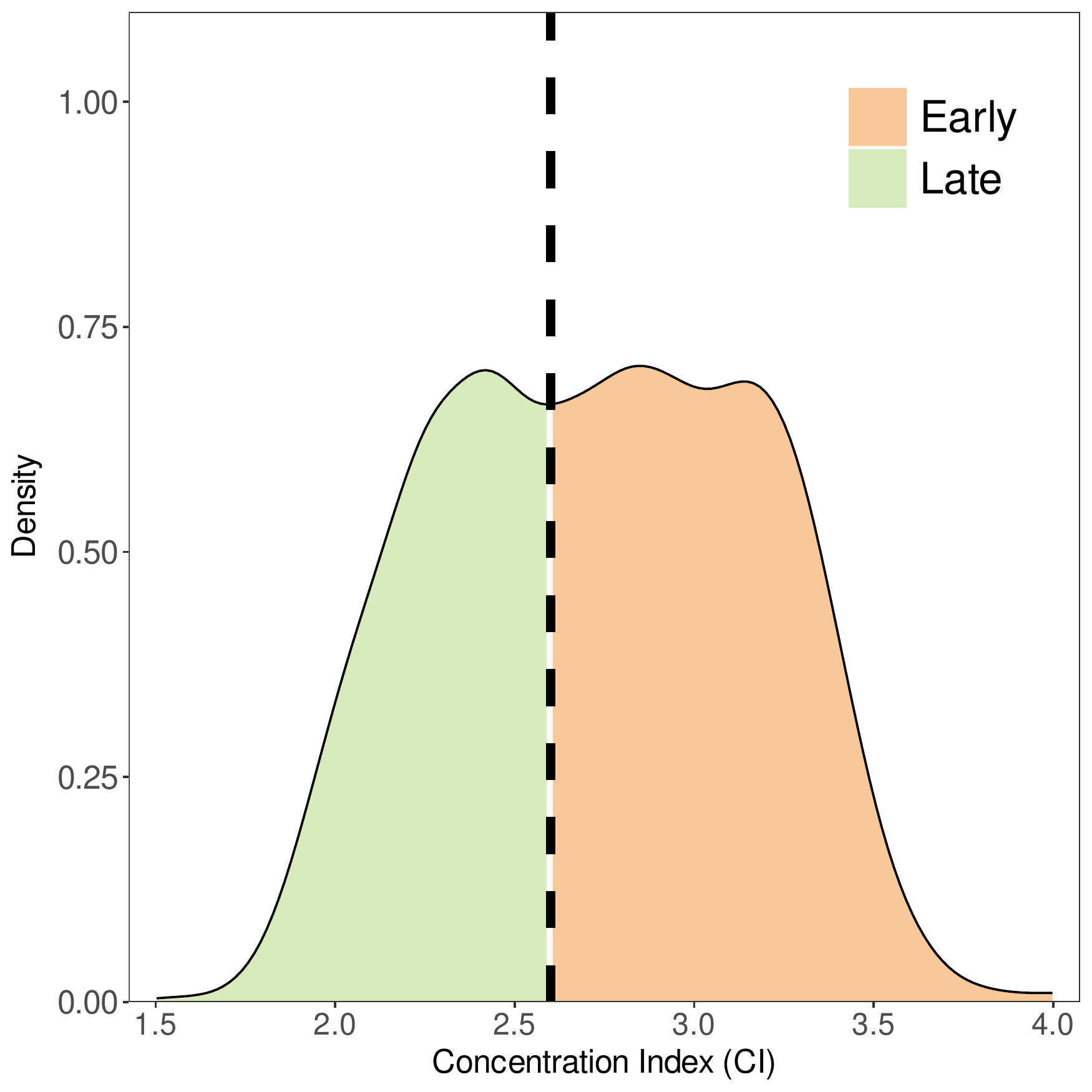}
\includegraphics[width=0.50\textwidth]{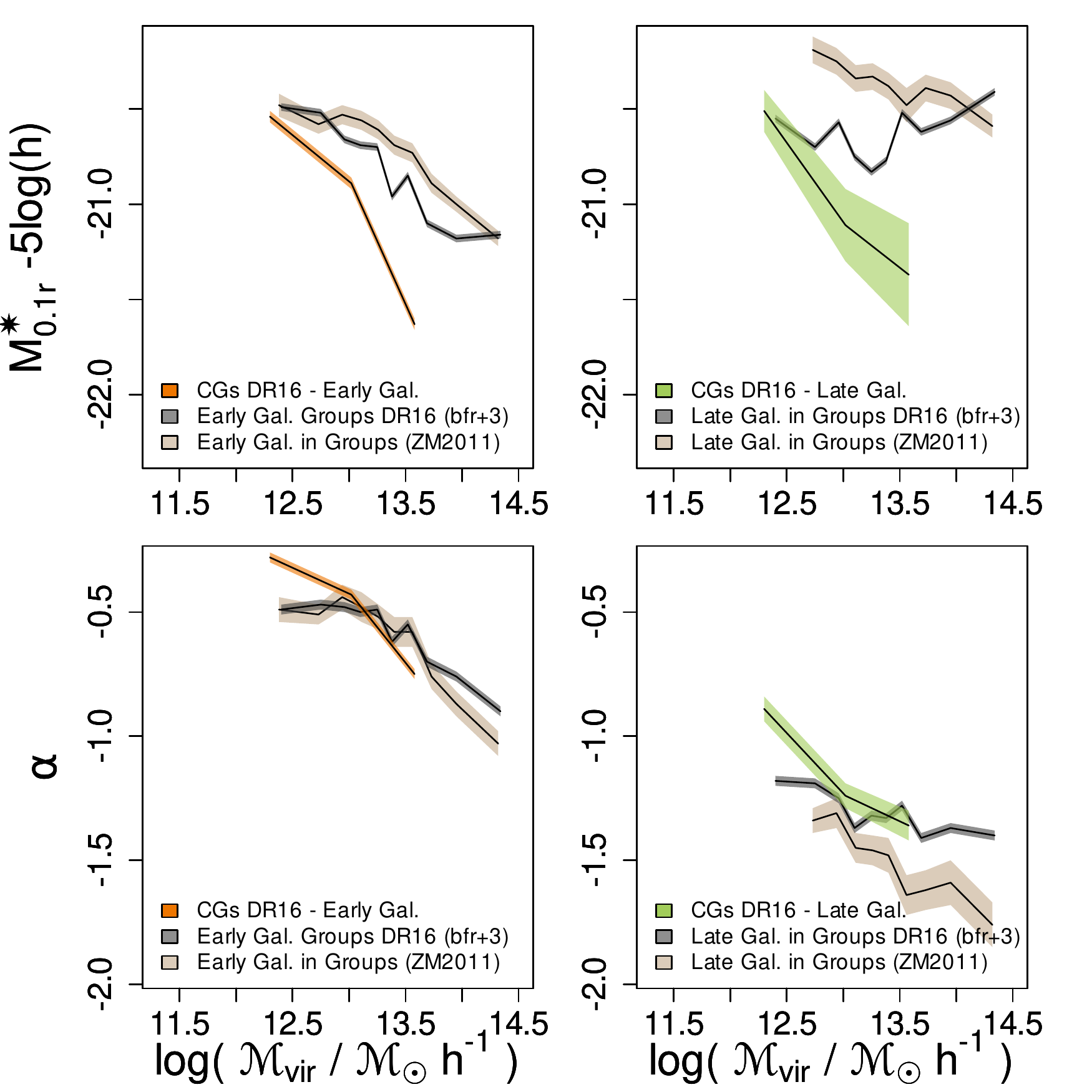}
\includegraphics[trim=0 -2cm 0 0, width=0.40\textwidth]{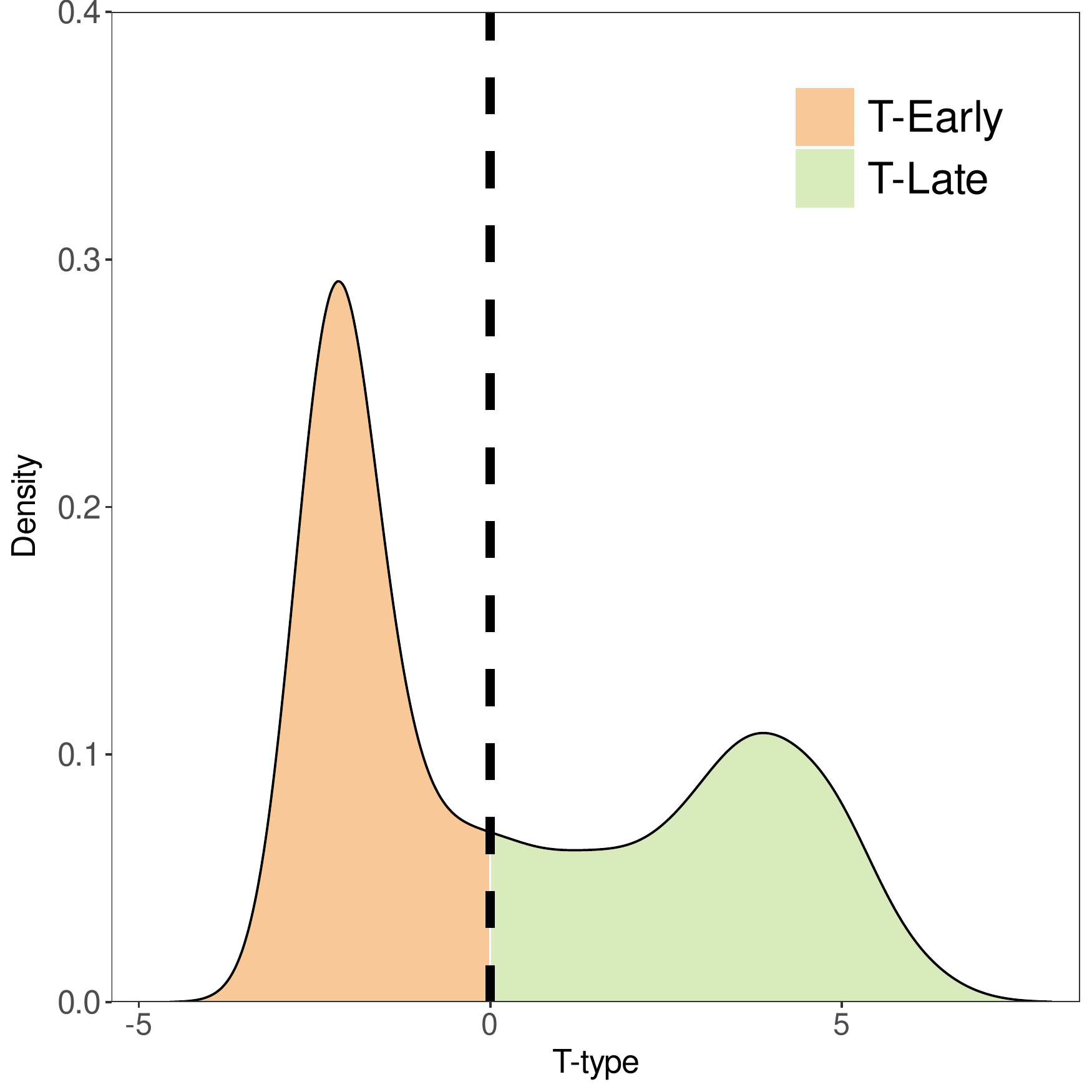}
\includegraphics[width=0.50\textwidth]{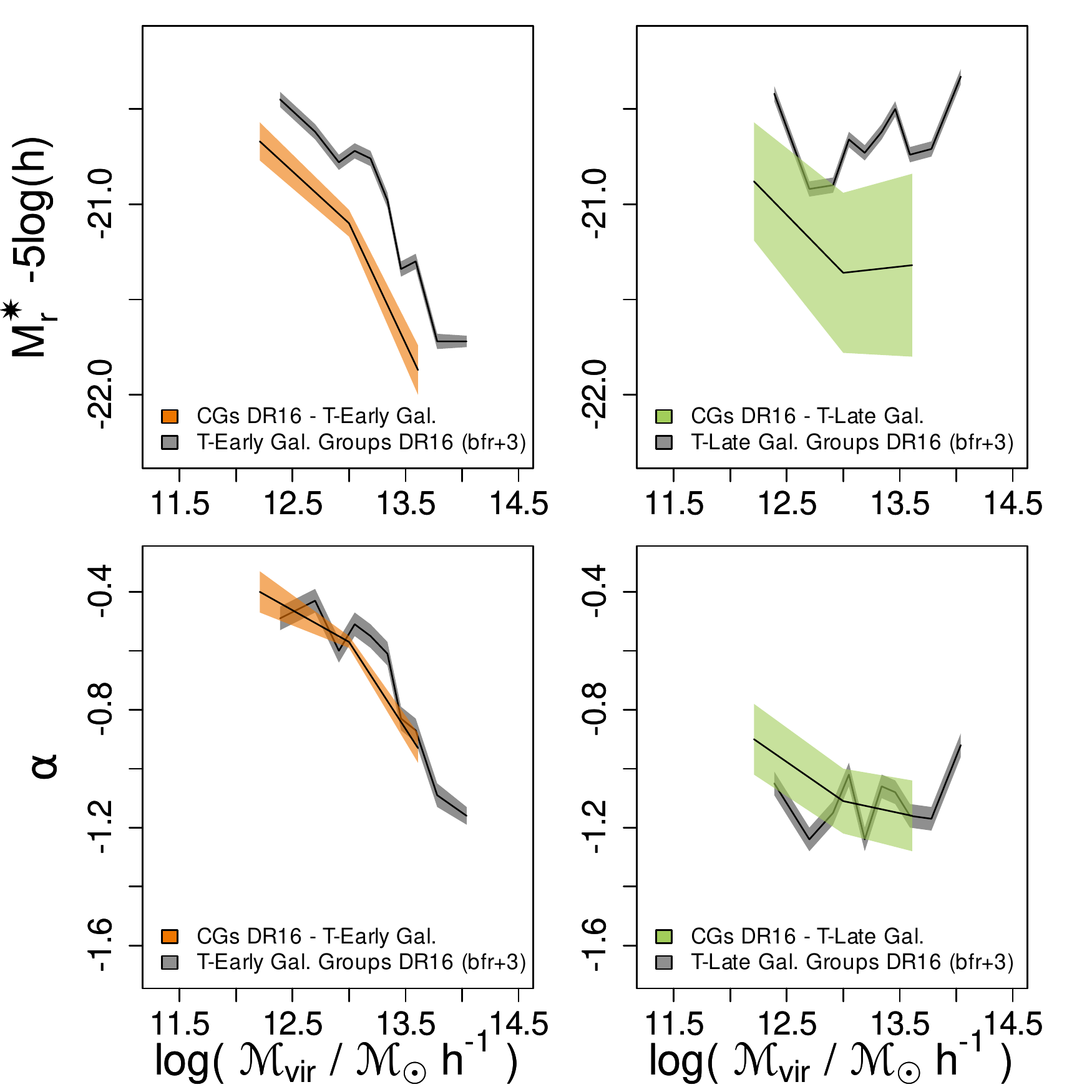}
\caption{The mass dependence of the LFs for the Late and Early galaxy population in CGs. The left upper plot shows the concentration index (CI) distribution for CG galaxy members. The black dashed vertical line is the cut in CI (2.6) defined by \citealt{strateva2001} to split galaxies into Early (to the right) and Late (to the left) subsamples (see text for details). The four upper right panels show the STY best-fitting Schechter LF parameters obtained for the Early and Late CG galaxy subsamples as a function of CGs virial masses. 
The left lower panel shows the T-type distribution for CG galaxy members. The black dashed vertical line is the cut in T-type (0.0) used by \citealt{ttypecat} to split galaxies into T-Early (to the left) and T-Late (to the right) subsamples (see text for details). The four lower right panels show the STY best-fitting Schechter LF parameters obtained for the T-Early and T-Late CG galaxy subsamples as a function of CGs virial masses.
For comparison, we included the estimations obtained for similar subsamples of Early and Late galaxy populations in galaxy groups identified in the SDSS DR7 by \citetalias{ZM11} (when using CI) and our own estimates for loose groups in the SDSS DR16 with a bright first-ranked galaxy and with at least 3 galaxy members that are at most three magnitudes fainter than the brightest galaxy in the group (bfr+3).
}
\label{fig:morfo}
\end{figure*}

By definition, the concentration index (CI) is the ratio of the Petrosian radii enclosing 90 and 50 per cent of the galaxy flux. Following \cite{strateva2001}, $CI=2.6$ is the cutoff value to separate between Early ($CI$ above the cut) and Late ($CI$ below the cut) galaxy types. However, it is necessary to be careful when using the estimates of the $CI$ for galaxies with $r_{50}$ below 1.6 arcsec, because seeing effects could lead to unreliable estimates. Roughly 12\% of galaxies in CGs may suffer from this problem. Hence, to avoid discarding those galaxies, we adopted a random assignment of $CI$ as a function of galaxy luminosity following \citetalias{ZM11}\footnote{We selected all galaxies in our parent galaxy catalogue with $r\le 16$ (where $CI$ incompleteness is below 1\%), computed the distributions of the $CI$ for different luminosity bins and used them to randomly assign $CI$ values (see \citetalias{ZM11} for details).}. The final distribution of $CI$ for galaxies in CGs is shown in the left upper plot of Fig.\ref{fig:morfo}, where the vertical dotted line is the $CI$ cutoff suggested by \cite{strateva2001}.

Four right upper panels in Fig.\ref{fig:morfo} show the mass dependence of the Schechter parameters for the Early (orange) and Late (green) galaxies in CGs. The comparison with loose groups shows similar results to those observed in the previous section with the Red and Blue galaxy populations. Early and Late galaxies in CGs have very similar characteristic magnitudes, showing the same trends as a function of groups virial mass, while those populations in loose galaxy groups have a more distinctive behaviour (an invariant brightness in the Late galaxies and a brightening for Early types as a function of mass). Despite the general trend for each galaxy type, either Early or Late type galaxies in CGs are mostly brighter than their counterparts in loose galaxy groups. When analysing the faint end slope of the LFs, Early galaxies in CGs behave very similarly to the same population inhabiting loose groups. On the other hand, Late type galaxies in CGs display a progressive steepening as a function of mass while an invariant behaviour ($\alpha \sim -1.3$) is observed in loose galaxy groups.  

Using the CI is not the only parameter used to separate galaxy samples into their Early and Late populations. A different attempt can be using a very old recipe devised by \cite{ttypeorig} in which each galaxy type (e.g. Hubble types) is associated with a given number. This association is known as T-type morphological classification, and its importance lies in the fact that it allows a smooth transition from bulges to disks with a simple numerical scale. The distribution of this morphological parameter has a bimodal behaviour that allows a relatively clean separation between two galaxy populations: early and late. But, the drawback is that a morphological classification of a very large sample of galaxies is an incredibly time-consuming task. However, the rise of the matching learning algorithms has enabled the possibility of performing this classification on large galaxy surveys (e.g., \citealt{huertas11}).   
Particularly, the work of \cite{ttypecat}\footnote{\url{https://vizier.cfa.harvard.edu/viz-bin/VizieR?-source=J/MNRAS/476/3661}} provided a morphological catalogue for more than half a million galaxies in the SDSS DR7 \citep{sdssdr7} obtained by the implementation of deep learning algorithms. 
These algorithms were trained using the catalogue of \cite{nair10}, that performed a T-type classification to a sample of galaxies visually inspected. In their work, \cite{ttypecat} rescaled the T-type to range from -3 to 10, where 0 correspond to S0, negative values to early-type galaxies, and positive values to late-type galaxies. 
The catalogue comprises 670722 galaxies in the SDSS DR7 with apparent magnitudes in the range $14\le r\le17.77$. Recently, \cite{ttypecat2} released a new catalogue with morphological classification for MaNGA galaxies (Mapping Nearby Galaxies at Apache Point Observatory) in the SDSS DR17 \citep{manga}, adding a total of 10127 morphological determinations\footnote{\url{https://www.sdss.org/dr17/data_access/value-added-catalogs/?vac_id=manga-morphology-deep-learning-dr17-catalog}}. 
Therefore, we use these two samples of SDSS galaxies with T-type morphology to add this property to our samples of galaxies in CGs and in loose groups in the SDSS DR16. 
From the cross-correlation between catalogues and releases, we found T-types for roughly $80\%$ of galaxies in our CGs, and $86\%$ of galaxies in our loose groups. To deal with this incompleteness, we decided to use only the CGs and loose groups where all their galaxy members have assigned a T-type morphology. This restriction leads us to a subsample of only 630 CGs ($\sim 45\%$) with T-types for all of their members, while for loose groups labelled as bfr+3, we restrict the sample to only 1141 ($\sim 41\%$) that have T-types determined for all of their galaxies. In the lower left panel of Fig.\ref{fig:morfo}, we show the distribution of T-type morphologies for galaxies inhabiting CGs, which is clearly bimodal. For this work, we adopted as a T-type $\le 0$ (including S0) as the T-Early galaxy sample, while the T-Late sample has galaxies with T-types $>0$.  

Four right lower panels in Fig.\ref{fig:morfo} show the mass dependence of the Schechter parameters for the T-Early (orange) and T-Late (green) galaxies in CGs. Despite the larger uncertainties obtained for the parameters, the comparison with loose groups shows very similar results to those observed previously for the Early and Late populations using the CI. T-Early and T-Late galaxies in CGs have very similar characteristic magnitudes, proving to be brighter than those populations in loose galaxy groups. Regarding the faint end slope of the LFs, T-Early and T-Late galaxies in CGs behave very similarly to the same population in loose groups.

\begin{figure*}
\centering
\includegraphics[trim=0 -2cm 0 0, width=0.40\textwidth]{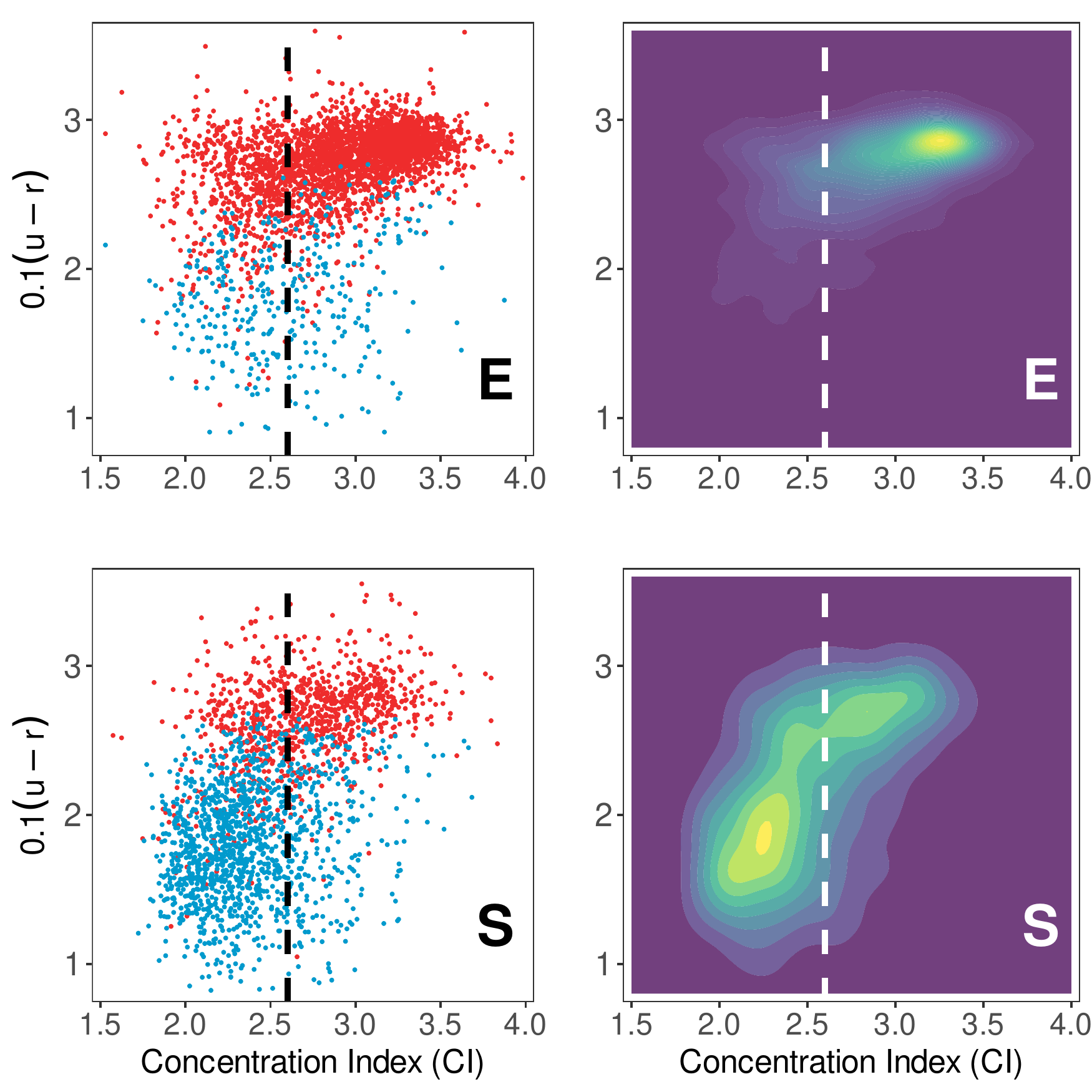}
\includegraphics[width=0.50\textwidth]{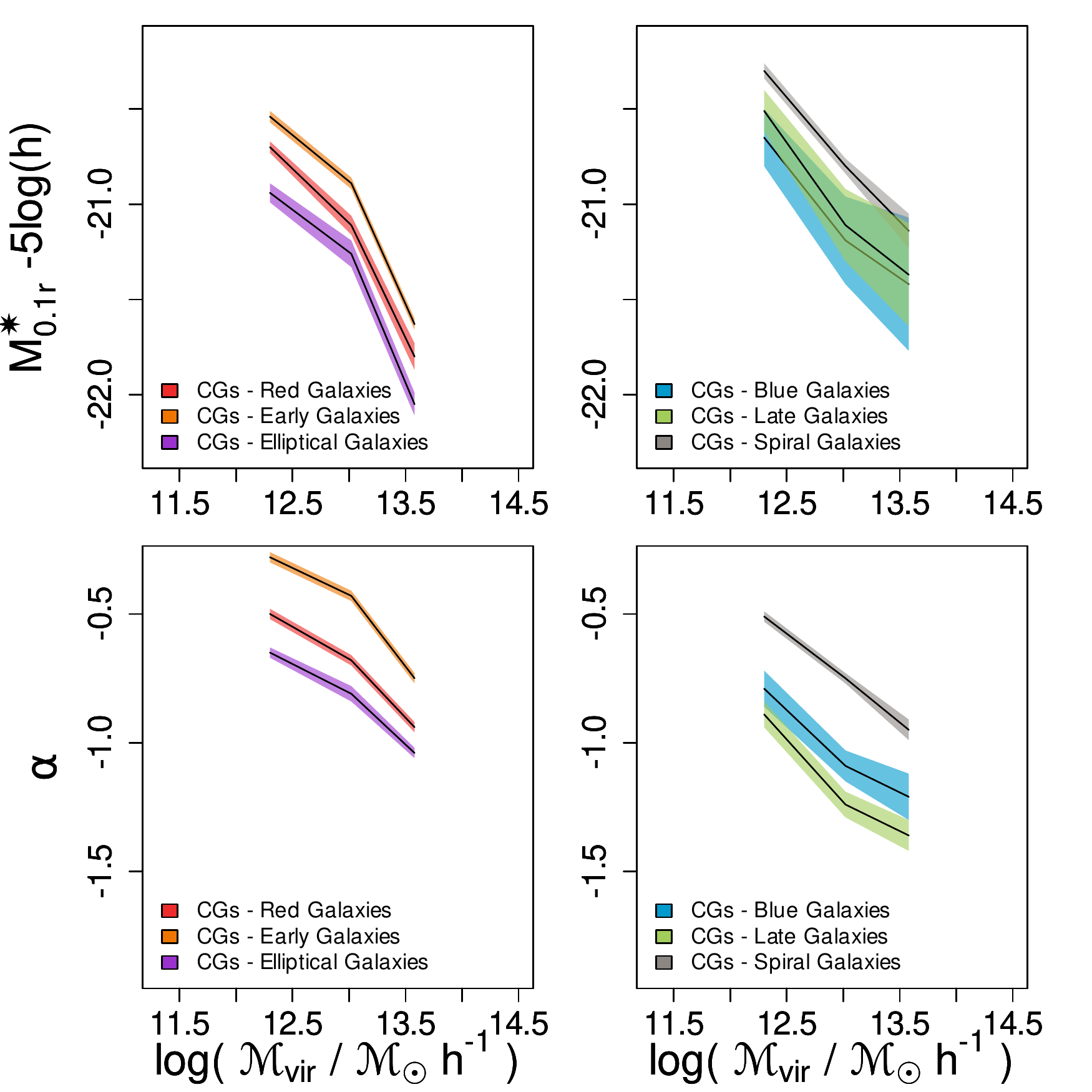}
\caption{The mass dependence of the LFs for Spiral and Elliptical galaxies in CGs. The four panels at the left show the colour-CI plane for CGs galaxy members. Red and Blue galaxy populations are drawn with their corresponding colours, while black vertical dashed line indicates the separation between Early and Late galaxy subsamples according to CI. Upper panels are constructed only for those CG galaxy members classified morphologically as Elliptical galaxies (E) according to the Galaxy Zoo database \citep{galaxyzoo1,galaxyzoo2}, while lower panels show the same but for those classified as Spiral galaxies (S). The four panels at the right show the STY best-fitting Schechter LF parameters obtained for the E and S CG galaxy subsamples as a function of CGs virial masses. For comparison, we included the previous determinations when CG galaxy populations are divided according to colour and CI.}
\label{fig:all}
\end{figure*}

\subsubsection{Elliptical and Spiral galaxies}
Using galaxy colours, concentration indexes, and T-type morphologies is an attempt to replace the lack of a morphological classification into Hubble galaxy types. Some authors have stated that morphology indicators (such as galaxy colour or CI) are less subjective than the traditional morphological classification in Hubble types, and more suitable to perform galaxy evolution and formation studies (e.g. \citealt{strateva2001}). However, the Hubble morphological types have always been of interest for astronomers of the last century. Historically, this type of classification was made by a visual inspection being an expensive and laborious task in terms of time invested. Therefore, performing this type of classification on the current large galaxy redshift surveys may seem not worth it.
Nevertheless, this type of titanic task was performed by what is known as the Galaxy Zoo Project\footnote{\url{www.galaxyzoo.org}}. A citizen science project that was launched in 2007, and embarked on the visual classification of approximately 1 million galaxies in the SDSS. The main sample of galaxies with their morphological classification (performed by more than $100 \, 000$ volunteers) was first released by \cite{galaxyzoo1} and \cite{galaxyzoo2}. This morphological galaxy information has allowed (for the scientific team of the project alone) the publication of more than sixty papers during the first 12 years of the project (see \citealt{galaxyzoorev} for a review). 

Hence, we complement our study by performing a new galaxy classification using the morphological types obtained from the Galaxy Zoo project based on the fraction of votes to be an Elliptical (E) and Spiral (S)\footnote{Actually, we used the fraction of votes for combined spiral, that include clockwise/anticlockwise spirals and edge-on disks.}. For each galaxy, we adopt the morphology with fraction of votes larger than 50\%. For those cases when all the fractions are below the threshold, we visually inspected the galaxy to decide which morphological type could be the most appropriate. We are aware that a threshold of 50\% entails between 15-20\% of misidentifications \citep{strateva2001}. Nevertheless, we take this morphological assignment as a first order approximation and, despite their limitations, we think that the possible misidentification of roughly a fifth of the galaxies should not substantially modify the trends shown by the remaining well-classified galaxies. It should also be taken into account that the CG sample is a low redshift sample, therefore the probability of misidentification could be lower.

Four left panels of Fig.\ref{fig:all} show the scatter plots and density maps in the colour vs. $CI$ plane for E and S galaxies in CGs. These plots show the interrelation among the different morphological indicators. For instance, Red/Early galaxies dominate the E subsample of galaxies, but there is also a small component of Blue/Late galaxies. When analysing the S galaxy subsample, a majority of Blue/Late galaxies make up the sample, but a non-negligible sample of Red/Early galaxies is also present. 

Four right panels of Fig.\ref{fig:all} show the comparison of the dependence of the Schechter parameters as a function of group mass for galaxies in CGs split using all the morphological indicators adopted in this work.  Firstly, all trends as a function of groups mass are present regardless the morphological indicator. However, the differences among the different indicators are related with the relative value of the Schechter parameters. The sample of E galaxies in CGs seems to be typically the brightest sample. Meanwhile, S galaxies in CGs are better described by the faintest values among the whole set of samples. 
From the comparison of the characteristic magnitude between Red vs. Blue, or Early vs. Late galaxies, those population are very alike to each other. Using the Hubble types instead, there is almost 1 magnitude difference between E and S galaxies in the high CG mass bin.  

When analysing the faint-end slope, E and S samples show similar trends, with S displaying a small deficiency of faint galaxies in comparison with the E population. 
Compared to the other indicators, E in CGs are less deficient in faint galaxies (more negative $\alpha$ values) compared with Red and Early samples, while S galaxies are the most deficient in faint population compared with their Blue and Late counterparts. 

\begin{figure}
\centering
\includegraphics[width=0.45\textwidth]{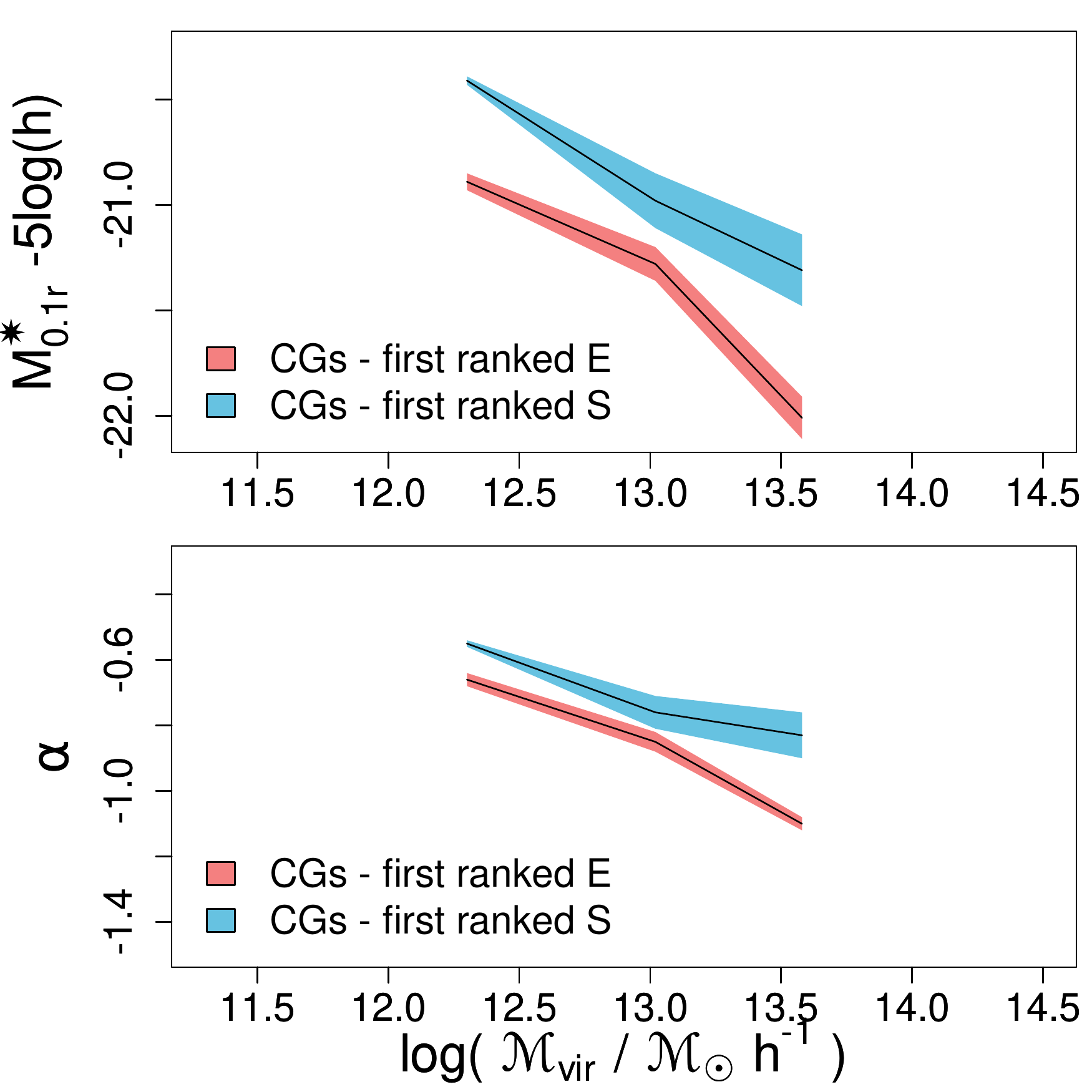}
\caption{The STY best-fitting Schechter LF parameters as a function of CG virial masses for CGs with E or S galaxies as its first ranked. Shaded areas are the semi-interquartile range for the median values of $\alpha$ and $M^{\ast}$ obtained from the markov chains.}
\label{fig:lf_mass_1st}
\end{figure}

\begin{figure}
\centering
\includegraphics[width=0.50\textwidth]{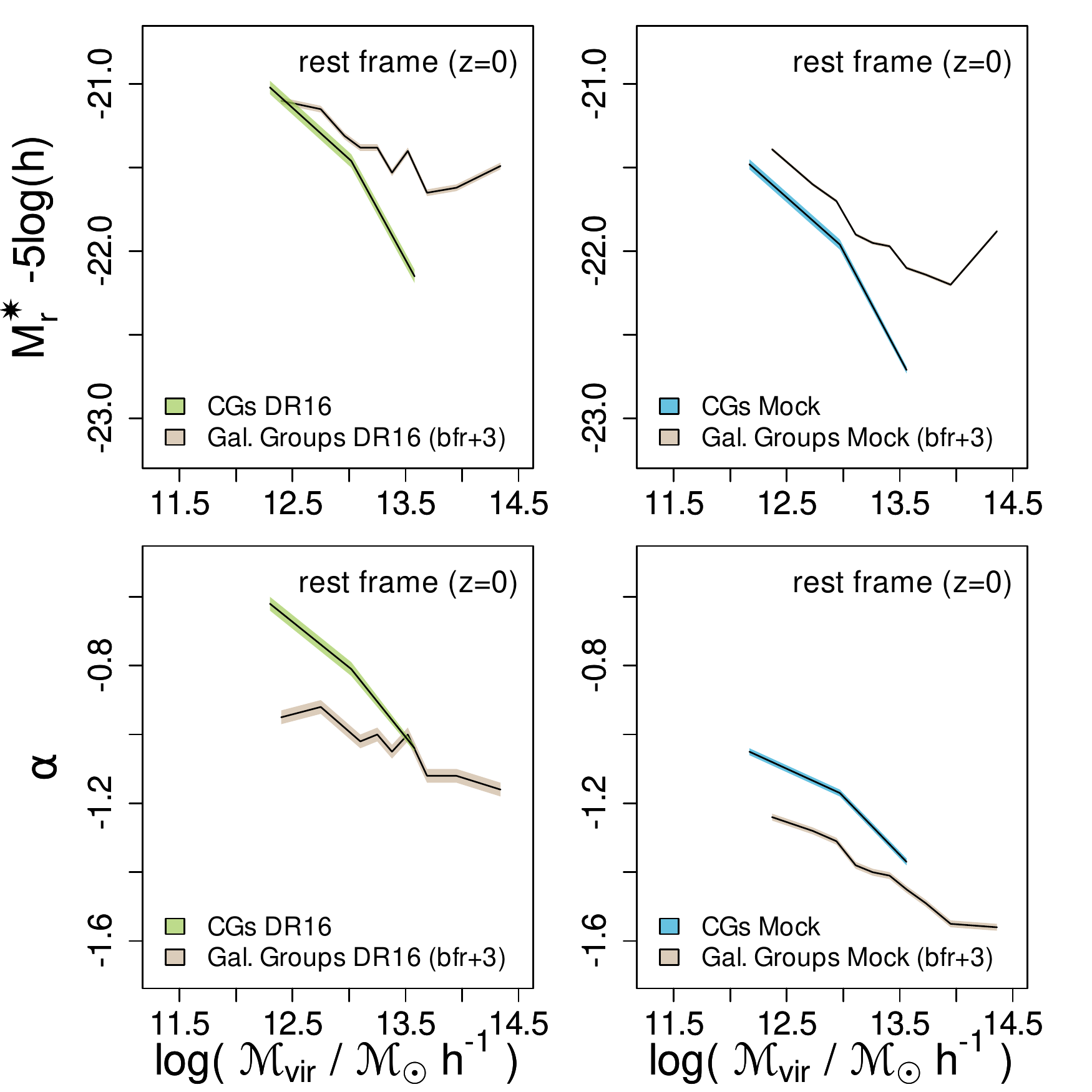}
\caption{The STY best-fitting Schechter LF parameters as a function of virial masses. Left panels show the comparison between CGs and loose galaxy groups in the SDSS DR16 using rest-frame absolute magnitudes in the r-band, while right panels show a similar comparison for galaxy systems identified in a mock galaxy catalogue. Shaded areas are the semi-interquartile range for the median values of $\alpha$ and $M^{\ast}$ obtained from the markov chains.}
\label{fig:lf_mass_mock}
\end{figure}

Finally, and following previous studies (e.g., \citealt{hunsberger98,Kelm&Focardi04}) in Fig.~\ref{fig:lf_mass_1st} we estimated the Schechter LF parameters as a function of mass for two CGs subsamples split according to the morphology of their first ranked galaxy. Our results show that CGs dominated by an E galaxy are characterised by brighter (up to $\sim 0.7$ mags for the most massive bin) galaxies than those observed in CGs dominated by an S galaxy. Also, a larger population of faint galaxies is observed in CGs dominated by E galaxies when compared with their counterparts dominated by S. Both results are observed in the whole range of virial masses, being much more noticeable for the most massive bin. 

\subsection{LF mass dependence in a mock catalogue}
The results obtained up to this point show distinctive behaviour of the galaxies that inhabit CGs. One question that naturally arises is whether currently existing semi-analytical models of galaxy formation (hereafter SAM) are capable of replicating such behaviour.
Therefore, as a supplementary case, we have performed a similar estimation of the LF Schechter parameters and its variation as a function of the virial mass for galaxies in CGs and loose systems identified in a particular mock galaxy lightcone. In Appendix~\ref{app:mock} we described the procedure to build a mock galaxy lightcone from the SAM of \cite{Henriques+20} that mimic the flux-limit of the SDSS DR16, as well as a complete description of the samples of CGs and loose galaxy groups identified in it.  

In Fig.~\ref{fig:lf_mass_mock} we show the comparison between the STY best-fitting Schecther LF parameters as a function of viral masses obtained for CGs and loose systems bfr+3 in the SDSS DR16 (left panels) and their counterparts identified in the mock catalogue (right panels). Given the limitations of our mock catalogue, the comparison is made using rest-frame absolute magnitudes at $z=0$ without evolution corrections. Despite this change in our usual procedure, it can be seen that the result obtained for the observational samples (left panels) does not differ much from that obtained previously (see Fig.~\ref{fig:lf_mass}) for the 0.1 shifted r-band (the only noticeable difference is an average $\sim 0.34$ brightening of the rest-frame characteristic absolute magnitudes). 

The dependence of the LF parameters as a function of mass is present in both, observations and mock catalogue. Although, a brightening of $0.5-0.6$ magnitudes in the $M^{\ast}$ of galaxies in the mock lightcone is observed (in both types of systems), and an excess of faint population ($\sim 0.37$ lower values of $\alpha$) in the mock catalogue respect to the observations.  
Anyway, this particular SAM is able to replicate the differences observed between galaxies in CGs and galaxies in loose groups. Galaxies in CGs are brighter (in terms of $M^{\ast}$) than galaxies in loose groups, regardless of the virial mass. At the same time, there is a lack of faint population in CGs in the whole range of virial masses (higher values of $\alpha$). 
 
\section{Summary and Discussions}
\label{sec:perspectives}
In this work, we analyse the possible influence of the Hickson-like CG environment on the luminosities of the galaxies that inhabit them. In particular, we studied the variation of the galaxy luminosities as a function of group virial masses. Our primary aim has been to compare the results obtained in Hickson-like CGs with those obtained in loose galaxy groups. 

To perform this task, we have constructed a new sample of Hickson-like CGs identified in an improved version of the SDSS DR16. We have followed a similar procedure as that developed by \cite{DiazGimenez+18}, but this time we have included triplets in our sample. Given the Hickson's magnitude concordance criterion there is a lack of a faint galaxy population, therefore, we have completed the CG samples by including all the galaxies fainter than 3 mags from the first ranked galaxy inside the isolation cylinder. 
We have also identified a sample of loose galaxy groups in the same parent galaxy sample using a FoF algorithm. Due to a restriction inherent to the Hickson criteria, which forces us to consider only those CGs with first ranked galaxy brighter than the apparent r-band magnitude limit of the SDSS minus 3 mags (flux limit criterion), we restricted the loose galaxy groups sample with the same criterion. We have also limited the sample of loose groups to those with at least three galaxy members within a three magnitude range from the brightest galaxy in order to mimic the CG concordance magnitude criterion. These restrictions allow us to perform a fair comparison, avoiding a possible luminosity bias among the samples.     

Our first result of an overall LF shows that galaxies in Hickson-like CGs have a characteristic absolute magnitude ($-21.31\pm0.02$) brighter than the observed for galaxies in the field or inhabiting loose groups. We also obtain a faint-end downward slope ($-0.87\pm0.01$), different to other environments.
A direct comparison with previous determinations for other CGs in the SDSS shows that our $M^{\ast}$ determination is consistent or little brighter ($\sim 0.1$ mags) than that obtained by \cite{coenda12}. On the other hand, our $\alpha$ estimate differs from the estimation of these authors, which is more in agreement with the expected for galaxies in the field. Our estimates for both, $M^{\ast}$ and $\alpha$, resemble the results obtained by \cite{zheng21} for their sample of embedded CGs.

Despite the difficulties in performing a fair comparison for different photometric bands as well as sampling issues, these findings show some common ground with other previous results in the literature. For instance, the high luminosity observed for galaxies in CGs was suggested 60 years ago by \citeauthor{limber60} \citep{hickson97}, while other works have reported that this was just a feature of the population of elliptical or early galaxies in CGs \citep{MendesdeOliveira&Hickson91,sulentic94,Kelm&Focardi04}.
Also, the deficiency of faint galaxies was also observed previously by \cite{ribeiro94} and \cite{barton96} (for CGs identified in a redshift survey), but we find that is not as extreme as the reported by \cite{MendesdeOliveira&Hickson91}.

To improve our understanding of these results, we studied the variation of the LF of galaxies in Hickson-like CGs as a function of group virial mass. The applied procedure is similar to that performed by \citetalias{ZM11} for loose groups in the SDSS DR7. We found a dependence of the LF Schechter parameters on the CGs masses. Moreover, we compared these results to those obtained from galaxies in loose groups. 
To perform a fair comparison, in this work we used our own sample of restricted loose groups identified in the SDSS DR16. We also observed a dependence of the LF parameters with group mass for galaxies in loose groups (as previously reported by \citetalias{ZM11}), but the variations are less pronounced than the observed in CGs. From the comparison between the galaxy luminosities in CGs and loose groups, we observed that galaxies in CGs more massive than $\sim 2\times 10^{13} \ {\rm {\cal M}_{\odot} \ h^{-1}} $ are the main responsible for the general brightening of the $M^{\ast}$ described above. 
On the other hand, the deficiency of faint galaxies in CGs observed for the whole sample can be attributed to the population of galaxies inhabiting low mass CGs.

We have also deepen our analysis by splitting the galaxy sample into different populations, using several morphology indicators. We observed that:
\begin{itemize}
    \item The LF parameters trends with mass for the Red galaxy populations in CGs and in loose groups are very similar to those observed for the total samples of galaxies. The only difference is that the trends for $\alpha$ are shifted to downward slopes for red galaxies in both, CGs and loose groups. 
    \item On the other hand, the luminosity trends with mass for the Blue galaxy populations show some differences depending on the type of system. While the Blue galaxies in loose groups show no apparent variation as a function of groups virial mass (already seen by \citetalias{ZM11}), their counterparts in CGs show a Blue population with variations as a function of mass similar to the observed for Red galaxies in CGS. 
    In loose groups, the values of $M^{\ast}$ for the Blue galaxies are fainter than the values for the Red galaxies, while in CGs the Blue galaxies are as bright as their Red companions.   
    The faint-end slope, $\alpha$, for Blue galaxies in CGs behaves similarly to the Red population but with a less notorious deficiency of faint galaxies.  
    \item Using either the concentration indexes or the T-type morphologies, we observed that the comparison between Early and Late galaxy populations in CGs and in loose groups is very similar to that described by the Red and Blue populations. 
    The trends with mass for $\alpha$ for the Early population in CGs and loose groups are shifted to more downward slopes than the Red galaxies. 
    This brightening of the Early population in CGs compared to Early galaxies in loose groups is in agreement with the results of \cite{Kelm&Focardi04}, but we also observed a potential brightening of the Late population that was not reported for those authors.
    \item When using a tentative assignment of Hubble types to split into Elliptical (E) and Spiral (S) galaxies, we observe that E in CGs show the brightest $M^{\ast}$ compared with Red and Early populations (in both, CGs and loose groups). 
    This result is in agreement with the findings of previous works \citep{MendesdeOliveira&Hickson91,sulentic94}. We also observed that S in CGs show the faintest $M^{\ast}$ compared with Blue and Late samples. 
    Analysing the faint-end slopes, we observed that E and S galaxies in CGs are similar in terms of the faint population (S galaxies are slightly more deficient in faint galaxies in comparison with the E). Noticeable differences emerge when compared to the other indicators: E galaxies are the less deficient in faints when compared with Red and Early samples, while S galaxies are the most deficient compared with Blue and Late types.
    \item Using the morphological classification through Hubble types, we also observed that CGs dominated by an E galaxy are characterised by galaxies typically brighter than their counterparts that inhabit CGs dominated by an S galaxy.
    Moreover, an excess of faint galaxies in CGs dominated by an E galaxy compared with the galaxies in CGs with an S first ranked galaxy is observed in agreement with \cite{hunsberger98}.
\end{itemize}

These findings suggest that properties of galaxies in CGs have undergone different processes or are driven by more extreme environmental effects than galaxies in loose systems. These conditions seem to affect, to a greater or lesser extent, all galaxy populations in CGs. 

Galaxy interactions and mergers emerge as plausible candidates for this differences, and as such, one may wonder what signs of these events are present in the sample analysed in this work.  

According to \cite{hickson97}, first-ranked galaxies should be elliptical if mergers were a dominant effect in CGs.
From the whole population of galaxies in CGs, we find that $\sim 60\%$ are E, while the percentage of CGs with a first ranked E galaxy is $\sim 66\%$. This difference in the percentages might indicate a small evidence in favour of galaxy mergers occurring inside CGs. 
In a previous study, \cite{zepf91} suggested that Blue-Elliptical galaxies might be an indication of recent mergers of gas-rich systems. We found that only $\sim 3\%$ of the first-ranked E galaxies are blue, while this percentage rise to $\sim 10\%$ when examining the whole sample of E galaxies in CGs. It means that this small fraction of the elliptical galaxies might show signs of recent star formation triggered by galaxy interactions. 

On the other hand, from numerical simulations, it has been shown that very gas-rich major mergers can produce spiral galaxies (e.g., \citealt{sparre17,martin18}). Also, numerical hydrodynamical simulations showed that ring and bars in galaxies are very likely formed due to galaxy interactions \citep{elagali18,peschken19}. In a recent work, \cite{guo20} reported that $\sim 70\%$ of their sample of Red-Spirals shows strong bars, inner/outer rings or signs of galaxy interactions/mergers. In our sample, we observe that $\sim 51\%$ of the CGs are dominated by a Red-Spiral first-ranked galaxy, while $\sim 42\%$ of the whole sample of S galaxies in CGs are red. Therefore, the considerable large fraction of Red-Spirals inhabiting CGs may suggest that galaxy interactions or mergers might be playing an important role in the galaxy evolution inside CGs.

Finally, we study if the currently available semi-analytical models of galaxy formation are able to reproduce some of these results. We used the SAM of galaxy formation constructed by \cite{Henriques+20} applied on top of the Millennium Simulation \citep{Springel+05}. From the comparison between mock and observations, we observed that, despite the mock galaxies are typically brighter and that the SAM produces more faint galaxies than in observations, the main features observed in the trends of the luminosity parameters in observations are replicated by the SAM. Therefore, it might indicate that the treatment in terms of dynamical evolution in this particular SAM is reasonable enough to reproduce these luminosity features as a function of group mass in both, CGs and loose systems.  
A more detailed study involving the different populations of mock galaxies would be the next step, although we know in advance that this will present more tension when compared to observations. By instance, \cite{Henriques+20} demonstrated that this particular SAM underestimate the fraction of bulge dominated galaxies and overestimate the fraction of disk dominated galaxies as a function of the stellar masses of the galaxies (see Fig. 17 therein).

Our results encourage further analysis of galaxies inhabiting CGs in order to disentangle the evolutionary history of galaxies in these particularly dense environments, one of the birthplaces of the brightest galaxies in the Universe.

\section*{Acknowledgements}
{\small  
The authors would like to thank the referee, for their thoughtful analysis and suggestions that helped us to improve the original manuscript. AZ and EDG dedicate this work to the memory of Gala, their tireless companion for the past 15 years. 

This publication uses as parent catalogue the SDSS Data Release 16 (DR16) which is one of the latest data releases of the SDSS-IV.
Funding for the Sloan Digital Sky 
Survey IV has been provided by the 
Alfred P. Sloan Foundation, the U.S. 
Department of Energy Office of 
Science, and the Participating 
Institutions. SDSS-IV acknowledges support and resources from the Center for High 
Performance Computing  at the 
University of Utah. The SDSS 
website is \url{www.sdss.org}.
SDSS-IV is managed by the 
Astrophysical Research Consortium 
for the Participating Institutions 
of the SDSS Collaboration including 
the Brazilian Participation Group, 
the Carnegie Institution for Science, 
Carnegie Mellon University, Center for 
Astrophysics | Harvard \& 
Smithsonian, the Chilean Participation 
Group, the French Participation Group, 
Instituto de Astrof\'isica de 
Canarias, The Johns Hopkins 
University, Kavli Institute for the 
Physics and Mathematics of the 
Universe (IPMU) / University of 
Tokyo, the Korean Participation Group, 
Lawrence Berkeley National Laboratory, 
Leibniz Institut f\"ur Astrophysik 
Potsdam (AIP),  Max-Planck-Institut 
f\"ur Astronomie (MPIA Heidelberg), 
Max-Planck-Institut f\"ur 
Astrophysik (MPA Garching), 
Max-Planck-Institut f\"ur 
Extraterrestrische Physik (MPE), 
National Astronomical Observatories of 
China, New Mexico State University, 
New York University, University of 
Notre Dame, Observat\'ario 
Nacional / MCTI, The Ohio State 
University, Pennsylvania State 
University, Shanghai 
Astronomical Observatory, United 
Kingdom Participation Group, 
Universidad Nacional Aut\'onoma 
de M\'exico, University of Arizona, 
University of Colorado Boulder, 
University of Oxford, University of 
Portsmouth, University of Utah, 
University of Virginia, University 
of Washington, University of 
Wisconsin, Vanderbilt University, 
and Yale University.

This publication also uses data generated via the Zooniverse.org platform, development of which is funded by generous support, including a Global Impact Award from Google, and by a grant from the Alfred P. Sloan Foundation.

The Millennium Simulation databases used in this paper and the web application providing online access to them were constructed as part of the activities of the German Astrophysical Virtual Observatory (GAVO).

This work has been partially supported by Consejo Nacional de Investigaciones Cient\'\i ficas y T\'ecnicas de la Rep\'ublica Argentina (CONICET) and the Secretar\'\i a de Ciencia y Tecnolog\'\i a de la Universidad de C\'ordoba (SeCyT)}

\section*{Data Availability} 
The main galaxy catalogue of the SDSS DR16 was downloaded from \url{https://skyserver.sdss.org/casjobs/}.
The sample to minimize the redshift incompleteness, mainly for bright galaxies, is a compiled sample of SDSS DR12 downloaded from \url{http://cosmodb.to.ee/}. 

The T-type morphological classification for SDSS galaxies was downloaded from \url{https://vizier.cfa.harvard.edu/viz-bin/VizieR?-source=J/MNRAS/476/3661} and \url{https://www.sdss.org/dr17/data_access/value-added-catalogs/?vac_id=manga-morphology-deep-learning-dr17-catalog}. Meanwhile, 
the morphology classification of SDSS galaxies performed by the Galaxy Zoo Project was downloaded from \url{http://www.galaxyzoo.org/}.

The simulated data used in this article were accessed from \url{http://gavo.mpa-garching.mpg.de/Millennium/}.

The final catalogue of CGs in the SDSS DR16 produced here will be made publicly available via the VizieR archive in \url{https://vizier.u-strasbg.fr/viz-bin/VizieR} after the publication of this work.

All the remaining derived data generated in this research will be shared on reasonable request to the corresponding authors.

\bibliography{biblio}

\begin{thebibliography}{}
\makeatletter
\relax
\def\mn@urlcharsother{\let\do\@makeother \do\$\do\&\do\#\do\^\do\_\do\%\do\~}
\def\mn@doi{\begingroup\mn@urlcharsother \@ifnextchar [ {\mn@doi@}
  {\mn@doi@[]}}
\def\mn@doi@[#1]#2{\def\@tempa{#1}\ifx\@tempa\@empty \href
  {http://dx.doi.org/#2} {doi:#2}\else \href {http://dx.doi.org/#2} {#1}\fi
  \endgroup}
\def\mn@eprint#1#2{\mn@eprint@#1:#2::\@nil}
\def\mn@eprint@arXiv#1{\href {http://arxiv.org/abs/#1} {{\tt arXiv:#1}}}
\def\mn@eprint@dblp#1{\href {http://dblp.uni-trier.de/rec/bibtex/#1.xml}
  {dblp:#1}}
\def\mn@eprint@#1:#2:#3:#4\@nil{\def\@tempa {#1}\def\@tempb {#2}\def\@tempc
  {#3}\ifx \@tempc \@empty \let \@tempc \@tempb \let \@tempb \@tempa \fi \ifx
  \@tempb \@empty \def\@tempb {arXiv}\fi \@ifundefined
  {mn@eprint@\@tempb}{\@tempb:\@tempc}{\expandafter \expandafter \csname
  mn@eprint@\@tempb\endcsname \expandafter{\@tempc}}}

\bibitem[\protect\citeauthoryear{{Abazajian} et~al.,}{{Abazajian}
  et~al.}{2009}]{sdssdr7}
{Abazajian} K.~N.,  et~al., 2009, \mn@doi [\apjs]
  {10.1088/0067-0049/182/2/543}, \href
  {https://ui.adsabs.harvard.edu/abs/2009ApJS..182..543A} {182, 543}

\bibitem[\protect\citeauthoryear{{Adelman-McCarthy} et~al.,}{{Adelman-McCarthy}
  et~al.}{2008}]{sdssdr6}
{Adelman-McCarthy} J.~K.,  et~al., 2008, \mn@doi [\apjs] {10.1086/524984},
  \href {https://ui.adsabs.harvard.edu/abs/2008ApJS..175..297A} {175, 297}

\bibitem[\protect\citeauthoryear{{Ahumada} et~al.,}{{Ahumada}
  et~al.}{2020}]{dr16}
{Ahumada} R.,  et~al., 2020, \mn@doi [\apjs] {10.3847/1538-4365/ab929e}, \href
  {https://ui.adsabs.harvard.edu/abs/2020ApJS..249....3A} {249, 3}

\bibitem[\protect\citeauthoryear{{Alam} et~al.,}{{Alam} et~al.}{2015}]{DR12b}
{Alam} S.,  et~al., 2015, \mn@doi [\apjs] {10.1088/0067-0049/219/1/12}, \href
  {http://adsabs.harvard.edu/abs/2015ApJS..219...12A} {219, 12}

\bibitem[\protect\citeauthoryear{{Baldry}, {Glazebrook}, {Brinkmann},
  {Ivezi{\'c}}, {Lupton}, {Nichol}  \& {Szalay}}{{Baldry}
  et~al.}{2004}]{baldry04}
{Baldry} I.~K.,  {Glazebrook} K.,  {Brinkmann} J.,  {Ivezi{\'c}} {\v{Z}}.,
  {Lupton} R.~H.,  {Nichol} R.~C.,   {Szalay} A.~S.,  2004, \mn@doi [\apj]
  {10.1086/380092}, \href
  {https://ui.adsabs.harvard.edu/abs/2004ApJ...600..681B} {600, 681}

\bibitem[\protect\citeauthoryear{{Barton}, {Geller}, {Ramella}, {Marzke}  \&
  {da Costa}}{{Barton} et~al.}{1996}]{barton96}
{Barton} E.,  {Geller} M.,  {Ramella} M.,  {Marzke} R.~O.,   {da Costa} L.~N.,
  1996, \mn@doi [\aj] {10.1086/118060}, \href
  {http://adsabs.harvard.edu/abs/1996AJ....112..871B} {112, 871}

\bibitem[\protect\citeauthoryear{{Beck}, {Dobos}, {Budav{\'a}ri}, {Szalay}  \&
  {Csabai}}{{Beck} et~al.}{2016}]{Beck+16}
{Beck} R.,  {Dobos} L.,  {Budav{\'a}ri} T.,  {Szalay} A.~S.,   {Csabai} I.,
  2016, \mn@doi [\mnras] {10.1093/mnras/stw1009}, \href
  {http://adsabs.harvard.edu/abs/2016MNRAS.460.1371B} {460, 1371}

\bibitem[\protect\citeauthoryear{{Beers}, {Flynn}  \& {Gebhardt}}{{Beers}
  et~al.}{1990}]{Beers+90}
{Beers} T.~C.,  {Flynn} K.,   {Gebhardt} K.,  1990, \mn@doi [\aj]
  {10.1086/115487}, \href {http://adsabs.harvard.edu/abs/1990AJ....100...32B}
  {100, 32}

\bibitem[\protect\citeauthoryear{{Blanton} \& {Roweis}}{{Blanton} \&
  {Roweis}}{2007}]{kcorrect}
{Blanton} M.~R.,  {Roweis} S.,  2007, \mn@doi [\aj] {10.1086/510127}, \href
  {https://ui.adsabs.harvard.edu/abs/2007AJ....133..734B} {133, 734}

\bibitem[\protect\citeauthoryear{{Blanton} et~al.,}{{Blanton}
  et~al.}{2003}]{blanton2003}
{Blanton} M.~R.,  et~al., 2003, \mn@doi [\apj] {10.1086/375776}, \href
  {https://ui.adsabs.harvard.edu/abs/2003ApJ...592..819B} {592, 819}

\bibitem[\protect\citeauthoryear{{Blanton} et~al.,}{{Blanton}
  et~al.}{2005}]{nyusdss}
{Blanton} M.~R.,  et~al., 2005, \mn@doi [\aj] {10.1086/429803}, \href
  {https://ui.adsabs.harvard.edu/abs/2005AJ....129.2562B} {129, 2562}

\bibitem[\protect\citeauthoryear{{Coenda}, {Muriel}  \&
  {Mart{\'\i}nez}}{{Coenda} et~al.}{2012}]{coenda12}
{Coenda} V.,  {Muriel} H.,   {Mart{\'\i}nez} H.~J.,  2012, \mn@doi [\aap]
  {10.1051/0004-6361/201118318}, \href
  {https://ui.adsabs.harvard.edu/abs/2012A&A...543A.119C} {543, A119}

\bibitem[\protect\citeauthoryear{{Colless} et~al.,}{{Colless}
  et~al.}{2001}]{2df1}
{Colless} M.,  et~al., 2001, \mn@doi [\mnras]
  {10.1046/j.1365-8711.2001.04902.x}, \href
  {http://adsabs.harvard.edu/abs/2001MNRAS.328.1039C} {328, 1039}

\bibitem[\protect\citeauthoryear{{Colless} et~al.,}{{Colless}
  et~al.}{2003}]{2df2}
{Colless} M.,  et~al., 2003, ArXiv Astrophysics e-prints, \href
  {http://adsabs.harvard.edu/abs/2003astro.ph..6581C} {}

\bibitem[\protect\citeauthoryear{{Corwin}, {Buta}  \& {de
  Vaucouleurs}}{{Corwin} et~al.}{1994}]{RC3b}
{Corwin} Jr. H.~G.,  {Buta} R.~J.,   {de Vaucouleurs} G.,  1994, \mn@doi [\aj]
  {10.1086/117225}, \href {http://adsabs.harvard.edu/abs/1994AJ....108.2128C}
  {108, 2128}

\bibitem[\protect\citeauthoryear{{D{\'{\i}}az-Gim{\'e}nez} \&
  {Mamon}}{{D{\'{\i}}az-Gim{\'e}nez} \& {Mamon}}{2010}]{DiazGimenez&Mamon10}
{D{\'{\i}}az-Gim{\'e}nez} E.,  {Mamon} G.~A.,  2010, \mnras, 409, 1227

\bibitem[\protect\citeauthoryear{{D{\'\i}az-Gim{\'e}nez}, {Zandivarez}  \&
  {Taverna}}{{D{\'\i}az-Gim{\'e}nez} et~al.}{2018}]{DiazGimenez+18}
{D{\'\i}az-Gim{\'e}nez} E.,  {Zandivarez} A.,   {Taverna} A.,  2018, \mn@doi
  [\aap] {10.1051/0004-6361/201833329}, \href
  {https://ui.adsabs.harvard.edu/abs/2018A&A...618A.157D} {618, A157}

\bibitem[\protect\citeauthoryear{{D{\'\i}az-Gim{\'e}nez}, {Taverna},
  {Zandivarez}  \& {Mamon}}{{D{\'\i}az-Gim{\'e}nez}
  et~al.}{2020}]{DiazGimenez+20}
{D{\'\i}az-Gim{\'e}nez} E.,  {Taverna} A.,  {Zandivarez} A.,   {Mamon} G.~A.,
  2020, \mn@doi [\mnras] {10.1093/mnras/stz3356}, \href
  {https://ui.adsabs.harvard.edu/abs/2020MNRAS.492.2588D} {492, 2588}

\bibitem[\protect\citeauthoryear{{Dom{\'\i}nguez S{\'a}nchez},
  {Huertas-Company}, {Bernardi}, {Tuccillo}  \& {Fischer}}{{Dom{\'\i}nguez
  S{\'a}nchez} et~al.}{2018}]{ttypecat}
{Dom{\'\i}nguez S{\'a}nchez} H.,  {Huertas-Company} M.,  {Bernardi} M.,
  {Tuccillo} D.,   {Fischer} J.~L.,  2018, \mn@doi [\mnras]
  {10.1093/mnras/sty338}, \href
  {https://ui.adsabs.harvard.edu/abs/2018MNRAS.476.3661D} {476, 3661}

\bibitem[\protect\citeauthoryear{{Dom{\'\i}nguez S{\'a}nchez}, {Margalef},
  {Bernardi}  \& {Huertas-Company}}{{Dom{\'\i}nguez S{\'a}nchez}
  et~al.}{2022}]{ttypecat2}
{Dom{\'\i}nguez S{\'a}nchez} H.,  {Margalef} B.,  {Bernardi} M.,
  {Huertas-Company} M.,  2022, \mn@doi [\mnras] {10.1093/mnras/stab3089}, \href
  {https://ui.adsabs.harvard.edu/abs/2022MNRAS.509.4024D} {509, 4024}

\bibitem[\protect\citeauthoryear{{Duarte} \& {Mamon}}{{Duarte} \&
  {Mamon}}{2014}]{duarte14}
{Duarte} M.,  {Mamon} G.~A.,  2014, \mn@doi [\mnras] {10.1093/mnras/stu378},
  \href {https://ui.adsabs.harvard.edu/abs/2014MNRAS.440.1763D} {440, 1763}

\bibitem[\protect\citeauthoryear{{Eisenstein} et~al.,}{{Eisenstein}
  et~al.}{2011}]{DR12a}
{Eisenstein} D.~J.,  et~al., 2011, \mn@doi [\aj] {10.1088/0004-6256/142/3/72},
  \href {http://adsabs.harvard.edu/abs/2011AJ....142...72E} {142, 72}

\bibitem[\protect\citeauthoryear{{Elagali}, {Lagos}, {Wong}, {Staveley-Smith},
  {Trayford}, {Schaller}, {Yuan}  \& {Abadi}}{{Elagali}
  et~al.}{2018}]{elagali18}
{Elagali} A.,  {Lagos} C. D.~P.,  {Wong} O.~I.,  {Staveley-Smith} L.,
  {Trayford} J.~W.,  {Schaller} M.,  {Yuan} T.,   {Abadi} M.~G.,  2018, \mn@doi
  [\mnras] {10.1093/mnras/sty2462}, \href
  {https://ui.adsabs.harvard.edu/abs/2018MNRAS.481.2951E} {481, 2951}

\bibitem[\protect\citeauthoryear{{Falco} et~al.,}{{Falco} et~al.}{1999}]{uzc}
{Falco} E.~E.,  et~al., 1999, \mn@doi [\pasp] {10.1086/316343}, \href
  {https://ui.adsabs.harvard.edu/abs/1999PASP..111..438F} {111, 438}

\bibitem[\protect\citeauthoryear{{Focardi} \& {Kelm}}{{Focardi} \&
  {Kelm}}{2002}]{Focardi&Kelm02}
{Focardi} P.,  {Kelm} B.,  2002, \mn@doi [\aap] {10.1051/0004-6361:20020377},
  \href {https://ui.adsabs.harvard.edu/abs/2002A&A...391...35F} {391, 35}

\bibitem[\protect\citeauthoryear{{Guo}, {Hao}, {Xia}, {Shi}, {Chen}, {Li}  \&
  {Gu}}{{Guo} et~al.}{2020}]{guo20}
{Guo} R.,  {Hao} C.-N.,  {Xia} X.,  {Shi} Y.,  {Chen} Y.,  {Li} S.,   {Gu} Q.,
  2020, \mn@doi [\apj] {10.3847/1538-4357/ab9b75}, \href
  {https://ui.adsabs.harvard.edu/abs/2020ApJ...897..162G} {897, 162}

\bibitem[\protect\citeauthoryear{Hastings}{Hastings}{1970}]{hastings}
Hastings W.~K.,  1970, \mn@doi [Biometrika] {10.1093/biomet/57.1.97}, 57, 97

\bibitem[\protect\citeauthoryear{{Henriques}, {Yates}, {Fu}, {Guo},
  {Kauffmann}, {Srisawat}, {Thomas}  \& {White}}{{Henriques}
  et~al.}{2020}]{Henriques+20}
{Henriques} B. M.~B.,  {Yates} R.~M.,  {Fu} J.,  {Guo} Q.,  {Kauffmann} G.,
  {Srisawat} C.,  {Thomas} P.~A.,   {White} S. D.~M.,  2020, \mn@doi [\mnras]
  {10.1093/mnras/stz3233}, \href
  {https://ui.adsabs.harvard.edu/abs/2020MNRAS.491.5795H} {491, 5795}

\bibitem[\protect\citeauthoryear{{Hickson}}{{Hickson}}{1982}]{Hickson82}
{Hickson} P.,  1982, \mn@doi [\apj] {10.1086/159838}, \href
  {http://adsabs.harvard.edu/abs/1982ApJ...255..382H} {255, 382}

\bibitem[\protect\citeauthoryear{{Hickson}}{{Hickson}}{1997}]{hickson97}
{Hickson} P.,  1997, \mn@doi [\araa] {10.1146/annurev.astro.35.1.357}, \href
  {https://ui.adsabs.harvard.edu/abs/1997ARA&A..35..357H} {35, 357}

\bibitem[\protect\citeauthoryear{{Hickson}, {Mendes de Oliveira}, {Huchra}  \&
  {Palumbo}}{{Hickson} et~al.}{1992}]{Hickson92}
{Hickson} P.,  {Mendes de Oliveira} C.,  {Huchra} J.~P.,   {Palumbo} G.~G.,
  1992, \mn@doi [\apj] {10.1086/171932}, \href
  {http://adsabs.harvard.edu/abs/1992ApJ...399..353H} {399, 353}

\bibitem[\protect\citeauthoryear{{Hubble}}{{Hubble}}{1936}]{hubble36}
{Hubble} E.~P.,  1936, {Realm of the Nebulae}

\bibitem[\protect\citeauthoryear{{Huchra} \& {Geller}}{{Huchra} \&
  {Geller}}{1982}]{huchra82}
{Huchra} J.~P.,  {Geller} M.~J.,  1982, \mn@doi [\apj] {10.1086/160000}, \href
  {https://ui.adsabs.harvard.edu/abs/1982ApJ...257..423H} {257, 423}

\bibitem[\protect\citeauthoryear{{Huchra} et~al.,}{{Huchra}
  et~al.}{2012}]{2mass3}
{Huchra} J.~P.,  et~al., 2012, \mn@doi [\apjs] {10.1088/0067-0049/199/2/26},
  \href {http://adsabs.harvard.edu/abs/2012ApJS..199...26H} {199, 26}

\bibitem[\protect\citeauthoryear{{Huertas-Company}, {Aguerri}, {Bernardi},
  {Mei}  \& {S{\'a}nchez Almeida}}{{Huertas-Company} et~al.}{2011}]{huertas11}
{Huertas-Company} M.,  {Aguerri} J.~A.~L.,  {Bernardi} M.,  {Mei} S.,
  {S{\'a}nchez Almeida} J.,  2011, \mn@doi [\aap]
  {10.1051/0004-6361/201015735}, \href
  {https://ui.adsabs.harvard.edu/abs/2011A&A...525A.157H} {525, A157}

\bibitem[\protect\citeauthoryear{{Humason}}{{Humason}}{1936}]{humason36}
{Humason} M.~L.,  1936, \mn@doi [\apj] {10.1086/143696}, \href
  {https://ui.adsabs.harvard.edu/abs/1936ApJ....83...10H} {83, 10}

\bibitem[\protect\citeauthoryear{{Hunsberger}, {Charlton}  \&
  {Zaritsky}}{{Hunsberger} et~al.}{1998}]{hunsberger98}
{Hunsberger} S.~D.,  {Charlton} J.~C.,   {Zaritsky} D.,  1998, \mn@doi [\apj]
  {10.1086/306201}, \href
  {https://ui.adsabs.harvard.edu/abs/1998ApJ...505..536H} {505, 536}

\bibitem[\protect\citeauthoryear{{Jarrett}, {Chester}, {Cutri}, {Schneider}  \&
  {Huchra}}{{Jarrett} et~al.}{2003}]{2mass1}
{Jarrett} T.~H.,  {Chester} T.,  {Cutri} R.,  {Schneider} S.~E.,   {Huchra}
  J.~P.,  2003, \mn@doi [\aj] {10.1086/345794}, \href
  {http://adsabs.harvard.edu/abs/2003AJ....125..525J} {125, 525}

\bibitem[\protect\citeauthoryear{{Kelm} \& {Focardi}}{{Kelm} \&
  {Focardi}}{2004}]{Kelm&Focardi04}
{Kelm} B.,  {Focardi} P.,  2004, \mn@doi [\aap] {10.1051/0004-6361:20040100},
  \href {http://adsabs.harvard.edu/abs/2004A%26A...418..937K} {418, 937}

\bibitem[\protect\citeauthoryear{{Krusch}, {Rosenbaum}, {Dettmar}, {Bomans},
  {Taylor}, {Aronica}  \& {Elwert}}{{Krusch} et~al.}{2006}]{krusch06}
{Krusch} E.,  {Rosenbaum} D.,  {Dettmar} R.~J.,  {Bomans} D.~J.,  {Taylor}
  C.~L.,  {Aronica} G.,   {Elwert} T.,  2006, \mn@doi [\aap]
  {10.1051/0004-6361:20054515}, \href
  {https://ui.adsabs.harvard.edu/abs/2006A&A...459..759K} {459, 759}

\bibitem[\protect\citeauthoryear{{Lange} et~al.,}{{Lange}
  et~al.}{2015}]{Lange+15}
{Lange} R.,  et~al., 2015, \mn@doi [\mnras] {10.1093/mnras/stu2467}, \href
  {http://adsabs.harvard.edu/abs/2015MNRAS.447.2603L} {447, 2603}

\bibitem[\protect\citeauthoryear{{Limber} \& {Mathews}}{{Limber} \&
  {Mathews}}{1960}]{limber60}
{Limber} D.~N.,  {Mathews} W.~G.,  1960, \mn@doi [\apj] {10.1086/146928}, \href
  {https://ui.adsabs.harvard.edu/abs/1960ApJ...132..286L} {132, 286}

\bibitem[\protect\citeauthoryear{{Lintott} et~al.,}{{Lintott}
  et~al.}{2011}]{galaxyzoo1}
{Lintott} C.,  et~al., 2011, \mn@doi [\mnras]
  {10.1111/j.1365-2966.2010.17432.x}, \href
  {https://ui.adsabs.harvard.edu/abs/2011MNRAS.410..166L} {410, 166}

\bibitem[\protect\citeauthoryear{{Lynden-Bell}}{{Lynden-Bell}}{1971}]{cmethod}
{Lynden-Bell} D.,  1971, \mn@doi [\mnras] {10.1093/mnras/155.1.95}, \href
  {https://ui.adsabs.harvard.edu/abs/1971MNRAS.155...95L} {155, 95}

\bibitem[\protect\citeauthoryear{{Mamon}, {Biviano}  \& {Bou{\'e}}}{{Mamon}
  et~al.}{2013}]{Mamon+13}
{Mamon} G.~A.,  {Biviano} A.,   {Bou{\'e}} G.,  2013, \mn@doi [\mnras]
  {10.1093/mnras/sts565}, \href
  {http://adsabs.harvard.edu/abs/2013MNRAS.429.3079M} {429, 3079}

\bibitem[\protect\citeauthoryear{{Martin}, {Kaviraj}, {Devriendt}, {Dubois}  \&
  {Pichon}}{{Martin} et~al.}{2018}]{martin18}
{Martin} G.,  {Kaviraj} S.,  {Devriendt} J.~E.~G.,  {Dubois} Y.,   {Pichon} C.,
   2018, \mn@doi [\mnras] {10.1093/mnras/sty1936}, \href
  {https://ui.adsabs.harvard.edu/abs/2018MNRAS.480.2266M} {480, 2266}

\bibitem[\protect\citeauthoryear{{Masters} \& {Galaxy Zoo Team}}{{Masters} \&
  {Galaxy Zoo Team}}{2020}]{galaxyzoorev}
{Masters} K.~L.,  {Galaxy Zoo Team} 2020, in {Valluri} M.,  {Sellwood} J.~A.,
  eds,  Proceedings of the International Astronomical Union Vol. 353, Galactic
  Dynamics in the Era of Large Surveys. pp 205--212 (\mn@eprint {arXiv}
  {1910.08177}), \mn@doi{10.1017/S1743921319008615}

\bibitem[\protect\citeauthoryear{{McConnachie}, {Patton}, {Ellison}  \&
  {Simard}}{{McConnachie} et~al.}{2009}]{McConnachie+09}
{McConnachie} A.~W.,  {Patton} D.~R.,  {Ellison} S.~L.,   {Simard} L.,  2009,
  \mn@doi [\mnras] {10.1111/j.1365-2966.2008.14340.x}, \href
  {http://adsabs.harvard.edu/abs/2009MNRAS.395..255M} {395, 255}

\bibitem[\protect\citeauthoryear{{Mendes de Oliveira} \& {Hickson}}{{Mendes de
  Oliveira} \& {Hickson}}{1991}]{MendesdeOliveira&Hickson91}
{Mendes de Oliveira} C.,  {Hickson} P.,  1991, \mn@doi [\apj] {10.1086/170559},
  \href {http://adsabs.harvard.edu/abs/1991ApJ...380...30M} {380, 30}

\bibitem[\protect\citeauthoryear{{Metropolis}, {Rosenbluth}, {Rosenbluth},
  {Teller}  \& {Teller}}{{Metropolis} et~al.}{1953}]{metropolis}
{Metropolis} N.,  {Rosenbluth} A.~W.,  {Rosenbluth} M.~N.,  {Teller} A.~H.,
  {Teller} E.,  1953, \mn@doi [\jcp] {10.1063/1.1699114}, \href
  {https://ui.adsabs.harvard.edu/abs/1953JChPh..21.1087M} {21, 1087}

\bibitem[\protect\citeauthoryear{{Morgan} \& {Mayall}}{{Morgan} \&
  {Mayall}}{1957}]{morgan57}
{Morgan} W.~W.,  {Mayall} N.~U.,  1957, \mn@doi [\pasp] {10.1086/127075}, \href
  {https://ui.adsabs.harvard.edu/abs/1957PASP...69..291M} {69, 291}

\bibitem[\protect\citeauthoryear{{Nair} \& {Abraham}}{{Nair} \&
  {Abraham}}{2010}]{nair10}
{Nair} P.~B.,  {Abraham} R.~G.,  2010, \mn@doi [\apjs]
  {10.1088/0067-0049/186/2/427}, \href
  {https://ui.adsabs.harvard.edu/abs/2010ApJS..186..427N} {186, 427}

\bibitem[\protect\citeauthoryear{{Peschken} \& {{\L}okas}}{{Peschken} \&
  {{\L}okas}}{2019}]{peschken19}
{Peschken} N.,  {{\L}okas} E.~L.,  2019, \mn@doi [\mnras]
  {10.1093/mnras/sty3277}, \href
  {https://ui.adsabs.harvard.edu/abs/2019MNRAS.483.2721P} {483, 2721}

\bibitem[\protect\citeauthoryear{{Planck Collaboration} et~al.,}{{Planck
  Collaboration} et~al.}{2014}]{Planck+14}
{Planck Collaboration} et~al., 2014, \mn@doi [\aap]
  {10.1051/0004-6361/201321591}, 571, A16

\bibitem[\protect\citeauthoryear{{Planck Collaboration} et~al.,}{{Planck
  Collaboration} et~al.}{2016}]{Planck+16}
{Planck Collaboration} et~al., 2016, \mn@doi [\aap]
  {10.1051/0004-6361/201525830}, \href
  {http://adsabs.harvard.edu/abs/2016A%26A...594A..13P} {594, A13}

\bibitem[\protect\citeauthoryear{{Ribeiro}, {de Carvalho}  \& {Zepf}}{{Ribeiro}
  et~al.}{1994}]{ribeiro94}
{Ribeiro} A.~L.~B.,  {de Carvalho} R.~R.,   {Zepf} S.~E.,  1994, \mn@doi
  [\mnras] {10.1093/mnras/267.1.L13}, \href
  {https://ui.adsabs.harvard.edu/abs/1994MNRAS.267L..13R} {267, L13}

\bibitem[\protect\citeauthoryear{{Robotham}, {Phillipps}  \& {de
  Propris}}{{Robotham} et~al.}{2010}]{robotham10}
{Robotham} A.,  {Phillipps} S.,   {de Propris} R.,  2010, \mn@doi [\mnras]
  {10.1111/j.1365-2966.2010.16252.x}, \href
  {https://ui.adsabs.harvard.edu/abs/2010MNRAS.403.1812R} {403, 1812}

\bibitem[\protect\citeauthoryear{{Sandage}, {Tammann}  \& {Yahil}}{{Sandage}
  et~al.}{1979}]{STY}
{Sandage} A.,  {Tammann} G.~A.,   {Yahil} A.,  1979, \mn@doi [\apj]
  {10.1086/157295}, \href
  {https://ui.adsabs.harvard.edu/abs/1979ApJ...232..352S} {232, 352}

\bibitem[\protect\citeauthoryear{{Schechter}}{{Schechter}}{1976}]{schechter}
{Schechter} P.,  1976, \mn@doi [\apj] {10.1086/154079}, \href
  {https://ui.adsabs.harvard.edu/abs/1976ApJ...203..297S} {203, 297}

\bibitem[\protect\citeauthoryear{{Shimasaku} et~al.,}{{Shimasaku}
  et~al.}{2001}]{shimasaku2001}
{Shimasaku} K.,  et~al., 2001, \mn@doi [\aj] {10.1086/322094}, \href
  {https://ui.adsabs.harvard.edu/abs/2001AJ....122.1238S} {122, 1238}

\bibitem[\protect\citeauthoryear{{Skrutskie} et~al.,}{{Skrutskie}
  et~al.}{2006}]{2mass2}
{Skrutskie} M.~F.,  et~al., 2006, \mn@doi [\aj] {10.1086/498708}, \href
  {http://adsabs.harvard.edu/abs/2006AJ....131.1163S} {131, 1163}

\bibitem[\protect\citeauthoryear{{Sparre} \& {Springel}}{{Sparre} \&
  {Springel}}{2017}]{sparre17}
{Sparre} M.,  {Springel} V.,  2017, \mn@doi [\mnras] {10.1093/mnras/stx1516},
  \href {https://ui.adsabs.harvard.edu/abs/2017MNRAS.470.3946S} {470, 3946}

\bibitem[\protect\citeauthoryear{{Springel} et~al.,}{{Springel}
  et~al.}{2005}]{Springel+05}
{Springel} V.,  et~al., 2005, \mn@doi [\nat] {10.1038/nature03597}, \href
  {http://adsabs.harvard.edu/abs/2005Natur.435..629S} {435, 629}

\bibitem[\protect\citeauthoryear{{Strateva} et~al.,}{{Strateva}
  et~al.}{2001}]{strateva2001}
{Strateva} I.,  et~al., 2001, \mn@doi [\aj] {10.1086/323301}, \href
  {https://ui.adsabs.harvard.edu/abs/2001AJ....122.1861S} {122, 1861}

\bibitem[\protect\citeauthoryear{{Sulentic} \& {Rabaca}}{{Sulentic} \&
  {Rabaca}}{1994}]{sulentic94}
{Sulentic} J.~W.,  {Rabaca} C.~R.,  1994, \mn@doi [\apj] {10.1086/174340},
  \href {https://ui.adsabs.harvard.edu/abs/1994ApJ...429..531S} {429, 531}

\bibitem[\protect\citeauthoryear{{Tempel}, {Tuvikene}, {Kipper}  \&
  {Libeskind}}{{Tempel} et~al.}{2017}]{tempel17}
{Tempel} E.,  {Tuvikene} T.,  {Kipper} R.,   {Libeskind} N.~I.,  2017, \mn@doi
  [\aap] {10.1051/0004-6361/201730499}, \href
  {http://adsabs.harvard.edu/abs/2017A%26A...602A.100T} {602, A100}

\bibitem[\protect\citeauthoryear{{Wake} et~al.,}{{Wake} et~al.}{2017}]{manga}
{Wake} D.~A.,  et~al., 2017, \mn@doi [\aj] {10.3847/1538-3881/aa7ecc}, \href
  {https://ui.adsabs.harvard.edu/abs/2017AJ....154...86W} {154, 86}

\bibitem[\protect\citeauthoryear{{Weinberg} \& {Kamionkowski}}{{Weinberg} \&
  {Kamionkowski}}{2003}]{weinberg+03}
{Weinberg} N.~N.,  {Kamionkowski} M.,  2003, \mn@doi [\mnras]
  {10.1046/j.1365-8711.2003.06421.x}, \href
  {http://adsabs.harvard.edu/abs/2003MNRAS.341..251W} {341, 251}

\bibitem[\protect\citeauthoryear{{Willett} et~al.,}{{Willett}
  et~al.}{2013}]{galaxyzoo2}
{Willett} K.~W.,  et~al., 2013, \mn@doi [\mnras] {10.1093/mnras/stt1458}, \href
  {https://ui.adsabs.harvard.edu/abs/2013MNRAS.435.2835W} {435, 2835}

\bibitem[\protect\citeauthoryear{{Willmer}}{{Willmer}}{1997}]{willmer97}
{Willmer} C.~N.~A.,  1997, \mn@doi [\aj] {10.1086/118522}, \href
  {https://ui.adsabs.harvard.edu/abs/1997AJ....114..898W} {114, 898}

\bibitem[\protect\citeauthoryear{{Yamanoi}, {Yagi}, {Komiyama}  \&
  {Koda}}{{Yamanoi} et~al.}{2020}]{yamanoi20}
{Yamanoi} H.,  {Yagi} M.,  {Komiyama} Y.,   {Koda} J.,  2020, \mn@doi [\aj]
  {10.3847/1538-3881/aba1ee}, \href
  {https://ui.adsabs.harvard.edu/abs/2020AJ....160...87Y} {160, 87}

\bibitem[\protect\citeauthoryear{{Zandivarez} \& {Mart{\'\i}nez}}{{Zandivarez}
  \& {Mart{\'\i}nez}}{2011}]{ZM11}
{Zandivarez} A.,  {Mart{\'\i}nez} H.~J.,  2011, \mn@doi [\mnras]
  {10.1111/j.1365-2966.2011.18878.x}, \href
  {https://ui.adsabs.harvard.edu/abs/2011MNRAS.415.2553Z} {415, 2553}

\bibitem[\protect\citeauthoryear{{Zandivarez}, {Mart{\'\i}nez}  \&
  {Merch{\'a}n}}{{Zandivarez} et~al.}{2006}]{zandivarez06}
{Zandivarez} A.,  {Mart{\'\i}nez} H.~J.,   {Merch{\'a}n} M.~E.,  2006, \mn@doi
  [\apj] {10.1086/503894}, \href
  {https://ui.adsabs.harvard.edu/abs/2006ApJ...650..137Z} {650, 137}

\bibitem[\protect\citeauthoryear{{Zepf} \& {Whitmore}}{{Zepf} \&
  {Whitmore}}{1991}]{zepf91}
{Zepf} S.~E.,  {Whitmore} B.~C.,  1991, \mn@doi [\apj] {10.1086/170811}, \href
  {https://ui.adsabs.harvard.edu/abs/1991ApJ...383..542Z} {383, 542}

\bibitem[\protect\citeauthoryear{{Zepf}, {de Carvalho}  \& {Ribeiro}}{{Zepf}
  et~al.}{1997}]{zepf97}
{Zepf} S.~E.,  {de Carvalho} R.~R.,   {Ribeiro} A. L.~B.,  1997, \mn@doi
  [\apjl] {10.1086/310914}, \href
  {https://ui.adsabs.harvard.edu/abs/1997ApJ...488L..11Z} {488, L11}

\bibitem[\protect\citeauthoryear{{Zheng} \& {Shen}}{{Zheng} \&
  {Shen}}{2020}]{zheng20}
{Zheng} Y.-L.,  {Shen} S.-Y.,  2020, \mn@doi [\apjs]
  {10.3847/1538-4365/ab5c26}, \href
  {https://ui.adsabs.harvard.edu/abs/2020ApJS..246...12Z} {246, 12}

\bibitem[\protect\citeauthoryear{{Zheng} \& {Shen}}{{Zheng} \&
  {Shen}}{2021}]{zheng21}
{Zheng} Y.-L.,  {Shen} S.-Y.,  2021, \mn@doi [\apj] {10.3847/1538-4357/abeaa2},
  \href {https://ui.adsabs.harvard.edu/abs/2021ApJ...911..105Z} {911, 105}

\bibitem[\protect\citeauthoryear{{de Vaucouleurs}}{{de
  Vaucouleurs}}{1961}]{devauc61}
{de Vaucouleurs} G.,  1961, \mn@doi [\apjs] {10.1086/190056}, \href
  {https://ui.adsabs.harvard.edu/abs/1961ApJS....5..233D} {5, 233}

\bibitem[\protect\citeauthoryear{{de Vaucouleurs}}{{de
  Vaucouleurs}}{1963}]{ttypeorig}
{de Vaucouleurs} G.,  1963, \mn@doi [\apjs] {10.1086/190084}, \href
  {https://ui.adsabs.harvard.edu/abs/1963ApJS....8...31D} {8, 31}

\bibitem[\protect\citeauthoryear{{de Vaucouleurs}, {de Vaucouleurs}, {Corwin},
  {Buta}, {Paturel}  \& {Fouqu{\'e}}}{{de Vaucouleurs} et~al.}{1991}]{RC3a}
{de Vaucouleurs} G.,  {de Vaucouleurs} A.,  {Corwin} Jr. H.~G.,  {Buta} R.~J.,
  {Paturel} G.,   {Fouqu{\'e}} P.,  1991, {Third Reference Catalogue of Bright
  Galaxies. Volume I: Explanations and references. Volume II: Data for galaxies
  between 0$^{h}$ and 12$^{h}$. Volume III: Data for galaxies between 12$^{h}$
  and 24$^{h}$.}

\makeatother
\end{thebibliography}

\appendix 
\section{Galaxy tables to complement the parent galaxy catalogue}

In this appendix, we show two lists of galaxies used in this work. In Table~\ref{tab:pofg} we quoted the 84 identification number (\texttt{objID}) of galaxies visually classified as Part of Galaxy.
In Table~\ref{tab:newz}, we detail the \texttt{objID} and redshifts of photometric galaxies around CG candidates. These redshifts are extracted from NASA/IPAC Extragalactic Database (NED). These new 159 redshifts were corrected to the CMB rest frame and added to the main catalogue.

\begin{table}
\setlength{\tabcolsep}{2pt}
\caption{SDSS \texttt{objID} of galaxies around potential CGs that were visually classified as Part of Galaxy (PofG). 
\label{tab:pofg}}
\begin{center}
\scriptsize
\begin{tabular}{rclrc}
\hline
\hline
&SDSSObjID && &SDSSObjID \\
\hline
\hline
   1&  1237671769072533603    &&    43 &  1237667910068863288\\
   2&  1237659162274103337    &&    44 &  1237671122692210867\\
   3&  1237657612339839115    &&    45 &  1237668292834689154\\
   4&  1237665128535752752    &&    46 &  1237668365851033830\\
   5&  1237665128535752748    &&    47 &  1237667549272146051\\
   6&  1237661212046524504    &&    48 &  1237668272980689000\\
   7&  1237658205059940491    &&    49 &  1237665532785459213\\
   8&  1237667912742338626    &&    50 &  1237665532785393884\\
   9&  1237668496321019995    &&    51 &  1237667255624990812\\
  10&  1237648720164421644    &&    52 &  1237661352167931944\\
  11&  1237651755080155195    &&    53 &  1237661352167997459\\
  12&  1237651192436293686    &&    54 &  1237668623008727174\\
  13&  1237671124303478913    &&    55 &  1237670964851966070\\
  14&  1237655498673422355    &&    56 &  1237668298741186706\\
  15&  1237655465923182607    &&    57 &  1237661386540187791\\
  16&  1237655499737726994    &&    58 &  1237661813886550028\\
  17&  1237655504034070593    &&    59 &  1237659343736406091\\
  18&  1237657874868535397    &&    60 &  1237659343736406092\\
  19&  1237658203421868039    &&    61 &  1237661971724828697\\
  20&  1237661950790467700    &&    62 &  1237659131678949555\\
  21&  1237658302205853770    &&    63 &  1237655470213103633\\
  22&  1237657606425215170    &&    64 &  1237658613595111439\\
  23&  1237654030328791130    &&    65 &  1237671958590455869\\
  24&  1237658630770720857    &&    66 &  1237665226237083742\\
  25&  1237658492281880652    &&    67 &  1237651736313659467\\
  26&  1237658613587181631    &&    68 &  1237648721761992788\\
  27&  1237661357538541706    &&    69 &  1237665226766483526\\
  28&  1237661388675022933    &&    70 &  1237665226766483525\\
  29&  1237659162274562084    &&    71 &  1237668292835147793\\
  30&  1237655369828991048    &&    72 &  1237662236941877296\\
  31&  1237664131016360008    &&    73 &  1237661950260215832\\
  32&  1237661125078417436    &&    74 &  1237659131672133703\\
  33&  1237659161735004216    &&    75 &  1237659131672133701\\
  34&  1237662301377134718    &&    76 &  1237665016311709830\\
  35&  1237661951327666295    &&    77 &  1237667539082608758\\
  36&  1237661871866576916    &&    78 &  1237657402419839059\\
  37&  1237665226766549013    &&    79 &  1237662199356850441\\
  38&  1237665226766549017    &&    80 &  1237665566076305424\\
  39&  1237662263774806286    &&    81 &  1237660671431540772\\
  40&  1237667783386988592    &&    82 &  1237667253462892551\\
  41&  1237667254009200669    &&    83 &  1237657591395909655\\
  42&  1237661360222961749    &&    84 &  1237661416068546747\\
\hline
\end{tabular}
\end{center}
\end{table}


\begin{table}
\setlength{\tabcolsep}{1.7pt}
\renewcommand{\arraystretch}{1.0}
\caption{CMB rest frame redshifts for $159$ SDSS DR16 photometric galaxies extracted from NASA/IPAC Extragalactic Database (NED).
\label{tab:newz}}
\begin{center}
\scriptsize
\begin{tabular}{rccrcc}
\hline
\hline
& SDSSObjID & $z_{\rm CMB}$ & &  SDSSObjID & $z_{\rm CMB}$ \\
\hline
\hline
   1 & 1237660635983577503  &  0.044936  &    81 &  1237664854724378868 &  0.066549\\
   2 & 1237666226954174550  &  0.033246  &    82 &  1237664854724378979 &  0.066272\\
   3 & 1237661383306969147  &  0.050763  &    83 &  1237662661602967961 &  0.074243\\
   4 & 1237651737930629583  &  0.031374  &    84 &  1237662661603033370 &  0.065097\\
   5 & 1237667254005923867  &  0.046909  &    85 &  1237667210504896600 &  0.044329\\
   6 & 1237661387080728729  &  0.031581  &    86 &  1237662640121708561 &  0.033158\\
   7 & 1237657776071508443  &  0.016309  &    87 &  1237661949707485195 &  0.038542\\
   8 & 1237654382519910501  &  0.069649  &    88 &  1237667323259977878 &  0.027622\\
   9 & 1237651737906380967  &  0.078746  &    89 &  1237667444048658552 &  0.385256\\
  10 & 1237648721249435821  &  0.045263  &    90 &  1237667324334637095 &  0.077876\\
  11 & 1237654949450350762  &  0.046585  &    91 &  1237667444048658637 &  0.031098\\
  12 & 1237651272972239031  &  0.053815  &    92 &  1237661949200433475 &  0.038263\\
  13 & 1237651823781347376  &  0.090075  &    93 &  1237661849858670688 &  0.025579\\
  14 & 1237651754539417743  &  0.067183  &    94 &  1237667733966487761 &  0.022081\\
  15 & 1237651754539417759  &  0.068323  &    95 &  1237667551956173016 &  0.022110\\
  16 & 1237651754539417761  &  0.068323  &    96 &  1237665372258959581 &  0.134524\\
  17 & 1237658203429994611  &  0.053726  &    97 &  1237665429169242387 &  0.076147\\
  18 & 1237654880743063871  &  0.077558  &    98 &  1237665429169242405 &  0.063324\\
  19 & 1237651754543218795  &  0.031046  &    99 &  1237665429169242258 &  0.103083\\
  20 & 1237655123940606350  &  0.039565  &   100 &  1237665429169242333 &  0.066050\\
  21 & 1237654601029779708  &  0.066634  &   101 &  1237665429169307848 &  0.078028\\
  22 & 1237654604787744979  &  0.068764  &   102 &  1237665429169242163 &  0.083218\\
  23 & 1237654391104602364  &  0.065427  &   103 &  1237667255624990809 &  0.026019\\
  24 & 1237654604253233398  &  0.068561  &   104 &  1237667915416666200 &  0.023606\\
  25 & 1237654604246024342  &  0.007451  &   105 &  1237662263241211987 &  0.023384\\
  26 & 1237655106768207964  &  0.042845  &   106 &  1237662237481238670 &  0.052611\\
  27 & 1237655124465483924  &  0.041485  &   107 &  1237662263778017419 &  0.020227\\
  28 & 1237658493355491546  &  0.008058  &   108 &  1237665227312857285 &  0.003253\\
  29 & 1237657775541125380  &  0.008914  &   109 &  1237665548887261197 &  0.029284\\
  30 & 1237658204493709598  &  0.053900  &   110 &  1237665428645806161 &  0.035332\\
  31 & 1237662264850579567  &  0.065900  &   111 &  1237651754005233804 &  0.076240\\
  32 & 1237658303278481478  &  0.048829  &   112 &  1237662641197547570 &  0.062757\\
  33 & 1237658918530056387  &  0.044534  &   113 &  1237650796755419236 &  0.075481\\
  34 & 1237658800967712945  &  0.061799  &   114 &  1237650796755484681 &  0.075481\\
  35 & 1237658800967712966  &  0.061799  &   115 &  1237651822173618256 &  0.029903\\
  36 & 1237657773930053777  &  0.055482  &   116 &  1237661386540187790 &  0.034778\\
  37 & 1237658612519076024  &  0.033207  &   117 &  1237671932820258987 &  0.038344\\
  38 & 1237657770705289448  &  0.039998  &   118 &  1237651752929853500 &  0.040191\\
  39 & 1237659119871721603  &  0.033404  &   119 &  1237659118799028304 &  0.033787\\
  40 & 1237659119871787085  &  0.060718  &   120 &  1237658302202970154 &  0.004251\\
  41 & 1237658803114475563  &  0.029342  &   121 &  1237661813886550021 &  0.008577\\
  42 & 1237662238020468926  &  0.038648  &   122 &  1237648704042631184 &  0.024057\\
  43 & 1237661971191431388  &  0.094033  &   123 &  1237662262712664133 &  0.034339\\
  44 & 1237661971191431389  &  0.094033  &   124 &  1237662195071123469 &  0.009220\\
  45 & 1237664093440377139  &  0.641805  &   125 &  1237662501081055448 &  0.031650\\
  46 & 1237663915730337883  &  0.054168  &   126 &  1237674651003650159 &  0.021187\\
  47 & 1237664668424536174  &  0.028981  &   127 &  1237664289396555920 &  0.028173\\
  48 & 1237664668424536084  &  0.030415  &   128 &  1237671990269837369 &  0.087060\\
  49 & 1237671958594912353  &  0.076643  &   129 &  1237668292835147790 &  0.027666\\
  50 & 1237664093439852642  &  0.023391  &   130 &  1237661817636454420 &  0.030739\\
  51 & 1237657630604197997  &  0.065121  &   131 &  1237661387617730621 &  0.077211\\
  52 & 1237671763712934142  &  0.094024  &   132 &  1237651737382289429 &  0.026851\\
  53 & 1237671763712934148  &  0.085424  &   133 &  1237654948378771564 &  0.055079\\
  54 & 1237671936578093074  &  0.040953  &   134 &  1237655109448106092 &  0.004078\\
  55 & 1237671992949866527  &  0.039349  &   135 &  1237662199356850421 &  0.028625\\
  56 & 1237660962937045185  &  0.065953  &   136 &  1237659133817978967 &  0.076406\\
  57 & 1237658300056666654  &  0.092247  &   137 &  1237664880485662813 &  0.734784\\
  58 & 1237658300056666881  &  0.029790  &   138 &  1237664879949906005 &  0.027096\\
  59 & 1237660764834889938  &  0.045550  &   139 &  1237658492798238895 &  0.056553\\
  60 & 1237662263785685079  &  0.046775  &   140 &  1237655373036912766 &  0.031698\\
  61 & 1237655503498313770  &  0.086227  &   141 &  1237658492280242186 &  0.069449\\
  62 & 1237655503498314069  &  0.086097  &   142 &  1237658492280242187 &  0.069449\\
  63 & 1237660343930323083  &  0.065459  &   143 &  1237667324334768210 &  0.016589\\
  64 & 1237660343930323283  &  0.069290  &   144 &  1237655468598165543 &  0.282153\\
  65 & 1237659162274562080  &  0.033366  &   145 &  1237655108902453308 &  0.034864\\
  66 & 1237661874014585108  &  0.025682  &   146 &  1237659325495443493 &  0.066296\\
  67 & 1237662698110386427  &  0.070327  &   147 &  1237668495776219326 &  0.022131\\
  68 & 1237660413718364224  &  0.053426  &   148 &  1237667911138279447 &  0.032838\\
  69 & 1237661137421402376  &  0.019852  &   149 &  1237661416068546838 &  0.026486\\
  70 & 1237661137421402367  &  0.018175  &   150 &  1237659162276135058 &  0.072853\\
  71 & 1237657774468759659  &  0.089264  &   151 &  1237667736134615184 &  0.047166\\
  72 & 1237661850937655373  &  0.043688  &   152 &  1237667736134615185 &  0.080856\\
  73 & 1237662336796000514  &  0.057952  &   153 &  1237667212653297785 &  0.058832\\
  74 & 1237661386530750668  &  0.041926  &   154 &  1237665128535752751 &  0.050944\\
  75 & 1237665023836815485  &  0.040194  &   155 &  1237667253462892552 &  0.025975\\
  76 & 1237665023836815489  &  0.040404  &   156 &  1237665429169307868 &  0.078345\\
  77 & 1237664853104263197  &  0.018095  &   157 &  1237662269148496244 &  0.104929\\
  78 & 1237664854724313390  &  0.067550  &   158 &  1237662262712598912 &  0.100640\\
  79 & 1237664854724313188  &  0.063104  &   159 &  1237648721761927310 &  0.068127\\
  80 & 1237664854724378646  &  0.063339  &       &                      &          \\
\hline
\end{tabular}
\end{center}
\end{table}

\section{The catalogue of CGs in the SDSS DR16}
\label{app:catalogue}

We present the catalogue of compact groups built from the Sloan Digital Sky Server Data Release 16 \citep{dr16}. In Table \ref{tab:groups} we show the properties of 1582 CGs included in the catalogue, while in Table \ref{tab:members} we list the properties of CG members.

\begin{table*}
\caption{Compact groups identified in SDSS DR16. \label{tab:groups}}
\begin{center}
\begin{tabular}{cccccccccc}
\hline
\hline
\texttt{CG}id & $N$ & RA & Dec  & $z_{mean}$ & $z_{median}$ & $\Theta_{\rm G}$ & $\mu_r$  & $r_{\rm b}$ & Flag \\
 & & [deg] & [deg] & & &   [arcmin] & [$\rm mag \, arcsec^{-2}$]  & [mag]    &  \\
\hline
\hline
   1 & 2   &   112.006  &    41.919&  0.059476 & 0.059180   &  0.588 &   22.740  &    14.309 & 0\\
   2 & 3   &   112.042  &    40.055&  0.050552 & 0.050411   &  2.189 &   25.724  &    14.153 & 0\\
   3 & 4   &   114.842  &    45.103&  0.078483 & 0.078286   &  1.310 &   25.018  &    14.719 & 0\\
   4 & 3   &   114.978  &    49.320&  0.022140 & 0.021969   &  3.929 &   26.169  &    14.187 & 0\\
   5 & 3   &   116.581  &    44.832&  0.031789 & 0.031575   &  4.091 &   25.986  &    13.147 & 0\\
\vdots & \vdots & \vdots & \vdots & \vdots & \vdots & \vdots  \vdots & \vdots & \vdots & \vdots\\
1578 & 5  &    247.547   &   36.247 & 0.075867 & 0.075328  &   1.835  &  25.535  &    14.479 & 1\\
1579 & 5  &    250.332   &   13.424 & 0.050747 & 0.050810  &   2.065  &  24.858  &    13.764 & 1\\
1580 & 6  &    252.344   &   26.592 & 0.055271 & 0.054724  &   1.659  &  24.733  &    14.228 & 1\\
1581 & 3  &    252.405   &   38.849 & 0.062536 & 0.062448  &   2.049  &  26.103  &    14.617 & 1\\
1582 & 3  &    255.884   &   36.087 & 0.063372 & 0.062938  &   1.261  &  24.267  &    13.794 & 1\\
\hline
\multicolumn{10}{p{.75\textwidth}}{Notes. \texttt{CG}id: group ID, $N$: number of galaxy members, RA: group centre right ascension (J2000), Dec: group centre declination (J2000), $z_{mean}$: group mean CMB redshift, $z_{median}$: group median CMB redshift, $\Theta_{\rm G}$: angular diameter of the smallest circumscribed circle, $\mu_r$: r-band group surface brightness,
$r_{\rm b}$: r-band observer-frame model apparent magnitude of the group brightest galaxy, Flag: $0=$ clean groups, $1=$ potentially contaminated groups.
This table is available in electronic form.
}
\end{tabular}
\end{center}
\end{table*}

\begin{table*}
\caption{Galaxy members of compact groups identified in SDSS DR16.\label{tab:members}}
\begin{center}
\begin{tabular}{ccccccc}
\hline
\hline
\texttt{CG}id & $N_m$ & \texttt{gal}id & RA & Dec & Redshift  & $r$  \\
  & & & [deg] & [deg] &  & [mag] \\
\hline
\hline
1       & 1 & 1237663916797722855   &   112.003 &     41.919 &  0.058831    & 14.309 \\
1       & 2 & 1237663916797722856   &   112.018 &     41.924 &  0.060418    & 14.906 \\
1       & 3 & 1237663916797722857   &   111.995 &     41.915 &  0.059180    & 17.045 \\
2       & 1 & 1237663531326243079   &   112.071 &     40.084 &  0.049958    & 14.153 \\
2       & 2 & 1237663531326243082   &   112.054 &     40.088 &  0.051286    & 15.838 \\
2       & 3 & 1237663531326243427   &   112.013 &     40.026 &  0.050411    & 17.147 \\
3       & 1 & 1237663786878238968   &   114.844 &     45.118 &  0.079790    & 14.719 \\
3       & 2 & 1237663786878238970   &   114.840 &     45.124 &  0.077569    & 16.380 \\
3       & 3 & 1237663786878238862   &   114.850 &     45.082 &  0.078908    & 16.813 \\
3       & 4 & 1237663786878239171   &   114.819 &     45.088 &  0.077664    & 17.100 \\

\vdots & \vdots & \vdots & \vdots & \vdots & \vdots & \vdots \\
\hline
\multicolumn{7}{p{.65\textwidth}}{Notes: \texttt{CG}id: group ID, $N_m$: galaxy index, \texttt{gal}id: galaxy ID, RA: right ascension (J2000), Dec: declination (J2000),Redshift: CMB redshift, $r$: r-band observer-frame model apparent magnitudes corrected for extinction in the AB system. Galaxies of each group are ordered by their apparent magnitudes from brightest to faintest.
This table is available in electronic form.}
\end{tabular}
\end{center}
\end{table*}

\section{Schechter parameters for the luminosity function of galaxies in groups}
\label{app:lfparam}
Table~\ref{tab:lfparam} quotes the Schechter's best-fitted parameters for all the luminosity functions of galaxies in CGs displayed in this work.

\begin{figure}
    \centering
    \includegraphics[width=9cm]{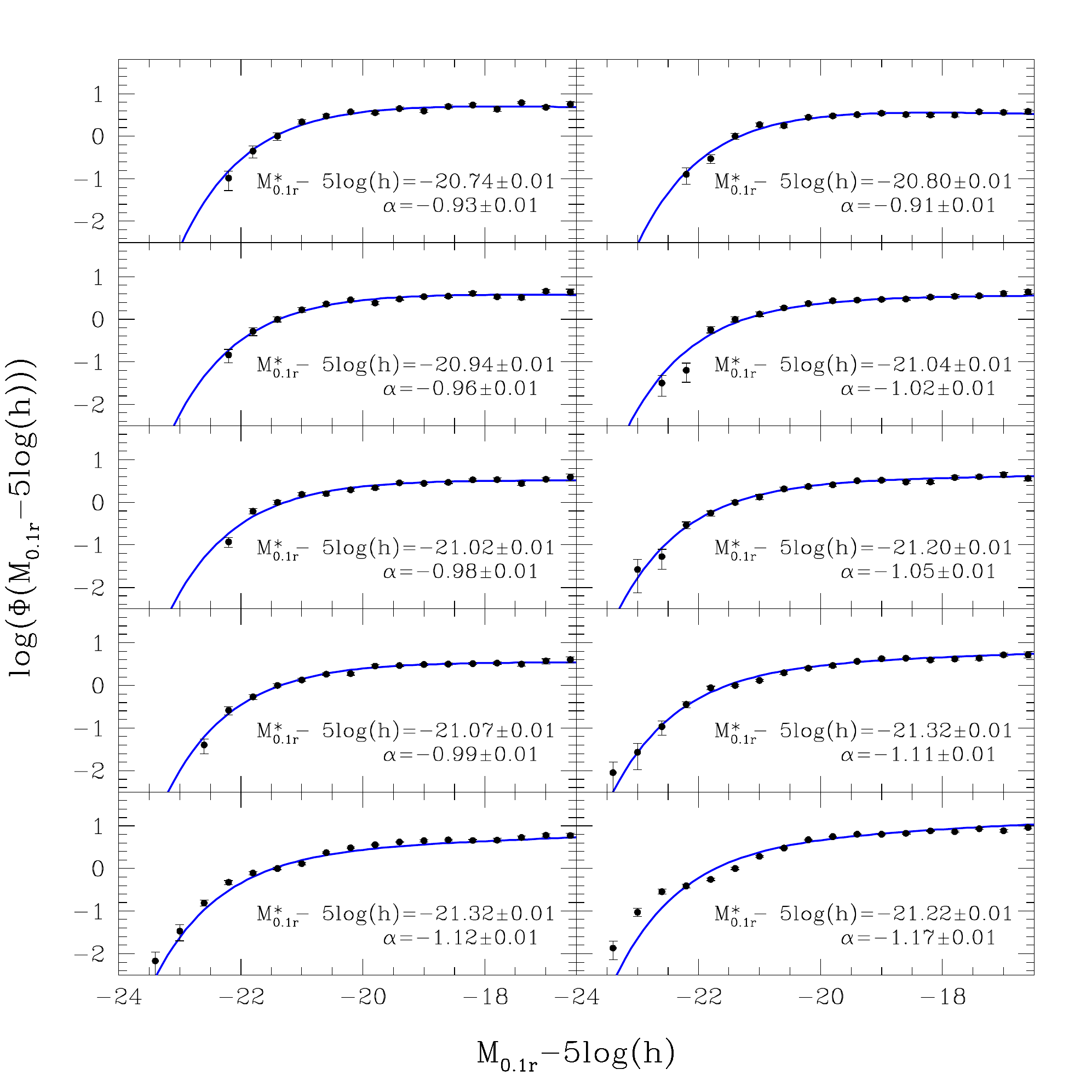}
    \caption{ The mass dependence of LFs of galaxies in SDSS DR16 galaxy groups subsample bfr+3. Points show the LFs estimated using the $C^{-}$ method (in arbitrary units), while error bars are estimated from a bootstrap resampling technique. 
    Solid lines show the STY best-fitting Schechter LF parameters obtained using 100 markov chains run in the likelihood space. These best-fitting parameters (see inset labels) where used to build Fig.\ref{fig:lf_mass}.}
    \label{fig:mass_fits}
\end{figure}

We also add in this section, as an example of the resulting Schechter LFs , Fig.\ref{fig:mass_fits} showing the fits obtained for the galaxy LF in SDSS DR16 galaxy groups (subsample bfr+3) for different group mass bins. As previously stated in the main text, in this work Schechter fits are a very good approximation of the general behaviour for the galaxy LFs.

\begin{table}
\setlength{\tabcolsep}{3pt}
\small
\centering
\caption{STY best-fitted LF parameters for galaxies in CGs. 
\label{tab:lfparam}}
\begin{tabular}{crrcc}
\hline
Mass bin & $N_{grp}$ & $N_{gal}$ & $M^{\ast}_{0.1r}-5\log(h)$ & $\alpha$ \\
\hline
\multicolumn{5}{c}{\bf Full galaxy sample}\\
\hline
12.30 & 470 & 1596 & $-20.71\pm0.04$ & $-0.63\pm0.02$ \\
13.02 & 470 & 1880 & $-21.13\pm0.05$ & $-0.81\pm0.02$ \\
13.58 & 472 & 2120 & $-21.79\pm0.07$ & $-1.03\pm0.02$ \\
\hline
\multicolumn{5}{c}{\bf Red galaxy sample}\\
\hline
12.30 & 470 & 1011 & $-20.70\pm0.03$ & $-0.50\pm0.02$ \\
13.02 & 470 & 1315 & $-21.11\pm0.05$ & $-0.68\pm0.02$ \\
13.58 & 472 & 1598 & $-21.80\pm0.07$ & $-0.94\pm0.02$ \\
\hline
\multicolumn{5}{c}{\bf Blue galaxy sample}\\
\hline
12.30 & 470 &  585 & $-20.65\pm0.15$ & $-0.79\pm0.07$ \\
13.02 & 470 &  565 & $-21.19\pm0.23$ & $-1.09\pm0.06$ \\
13.58 & 472 &  522 & $-21.42\pm0.35$ & $-1.21\pm0.09$ \\
\hline
\multicolumn{5}{c}{\bf Early galaxy sample}\\
\hline
12.30 & 470 &  934 & $-20.54\pm0.03$ & $-0.28\pm0.01$ \\
13.02 & 470 & 1126 & $-20.89\pm0.02$ & $-0.43\pm0.01$ \\
13.58 & 472 & 1314 & $-21.63\pm0.02$ & $-0.75\pm0.01$ \\
\hline
\multicolumn{5}{c}{\bf Late galaxy sample}\\
\hline
12.30 & 470 &  662 & $-20.51\pm0.11$ & $-0.89\pm0.05$ \\
13.02 & 470 &  754 & $-21.11\pm0.19$ & $-1.24\pm0.05$ \\
13.58 & 472 &  806 & $-21.37\pm0.27$ & $-1.36\pm0.06$ \\
\hline
\multicolumn{5}{c}{\bf T-Early galaxy sample}\\
\hline
12.21 & 209 & 322 & $-20.67\pm0.10$ & $-0.40\pm0.07$ \\
13.00 & 210 & 388 & $-21.10\pm0.07$ & $-0.57\pm0.02$ \\
13.61 & 211 & 506 & $-21.87\pm0.13$ & $-0.93\pm0.05$ \\
\hline
\multicolumn{5}{c}{\bf T-Late galaxy sample}\\
\hline
12.21 & 209 & 360 & $-20.88\pm0.31$ & $-0.90\pm0.12$ \\
13.00 & 210 & 377 & $-21.36\pm0.42$ & $-1.11\pm0.11$ \\
13.61 & 211 & 324 & $-21.32\pm0.48$ & $-1.16\pm0.12$ \\
\hline
\multicolumn{5}{c}{\bf Elliptical (E) galaxy sample}\\
\hline
12.30 & 470 &  808 & $-20.94\pm0.05$ & $-0.65\pm0.02$ \\
13.02 & 470 & 1112 & $-21.26\pm0.07$ & $-0.81\pm0.03$ \\
13.58 & 472 & 1403 & $-22.05\pm0.06$ & $-1.04\pm0.02$ \\
\hline
\multicolumn{5}{c}{\bf Spiral (S) galaxy sample}\\
\hline
12.30 & 470 &  788 & $-20.30\pm0.04$ & $-0.51\pm0.02$ \\
13.02 & 470 &  768 & $-20.80\pm0.04$ & $-0.75\pm0.02$ \\
13.58 & 472 &  717 & $-21.14\pm0.09$ & $-0.95\pm0.04$ \\
\hline
\multicolumn{5}{c}{\bf Gal. in CGs with first ranked E}\\
\hline
12.30 & 258 &  892 & $-20.89\pm0.04$ & $-0.66\pm0.02$ \\
13.02 & 307 & 1246 & $-21.28\pm0.08$ & $-0.85\pm0.03$ \\
13.58 & 358 & 1682 & $-22.01\pm0.10$ & $-1.10\pm0.02$ \\
\hline
\multicolumn{5}{c}{\bf Gal. in CGs with first ranked S}\\
\hline
12.30 & 212 &  704 & $-20.41\pm0.02$ & $-0.55\pm0.01$ \\
13.02 & 163 &  634 & $-20.98\pm0.13$ & $-0.76\pm0.05$ \\
13.58 & 114 &  438 & $-21.31\pm0.17$ & $-0.83\pm0.07$ \\
\hline
\hline
\end{tabular}
\parbox{\hsize}{\noindent Notes: 
Mass bin is in $\log({\cal M}/({\rm h^{-1}}{\cal M}_{\odot}))$ units; $N_{grp}$: number of CGs; $N_{gal}$: number of galaxies in CGs in the r-band absolute magnitudes range $[-24,-16]$.\\
}
\normalsize
\end{table}

\section{Mock galaxy lightcone: compact and loose groups}
\label{app:mock}
We use the publicly available galaxies produced by the semi-analytical model of galaxy formation (SAM) developed by \cite{Henriques+20} run on the Millennium Simulation \citep{Springel+05} and re-scaled to the Planck cosmology \citep{Planck+16}.  
We place an observer in one of the vertices of the simulation and produce an all-sky lightcone following the procedure described in previous works \citep{DiazGimenez+18, DiazGimenez+20}. Briefly, galaxies are extracted from the latest 16 outputs of the simulation to consider the evolution of properties and structures.  Redshifts of galaxies are computed using the comoving positions in the boxes and the peculiar velocities of each galaxy.
Observer frame apparent magnitudes are computed from the absolute magnitudes (interpolated between different snapshots) provided by the SAM and a k-decorrection is included following \cite{DiazGimenez+18} 
An observer frame apparent magnitude limit in the r-band of $17.77$ is adopted to select galaxies in the lightcone. The all-sky lightcone comprises $\sim 3\,590\,000$ galaxies. 

\subsection{Compact Groups}
\begin{figure}
    \centering
    \includegraphics[width=8cm]{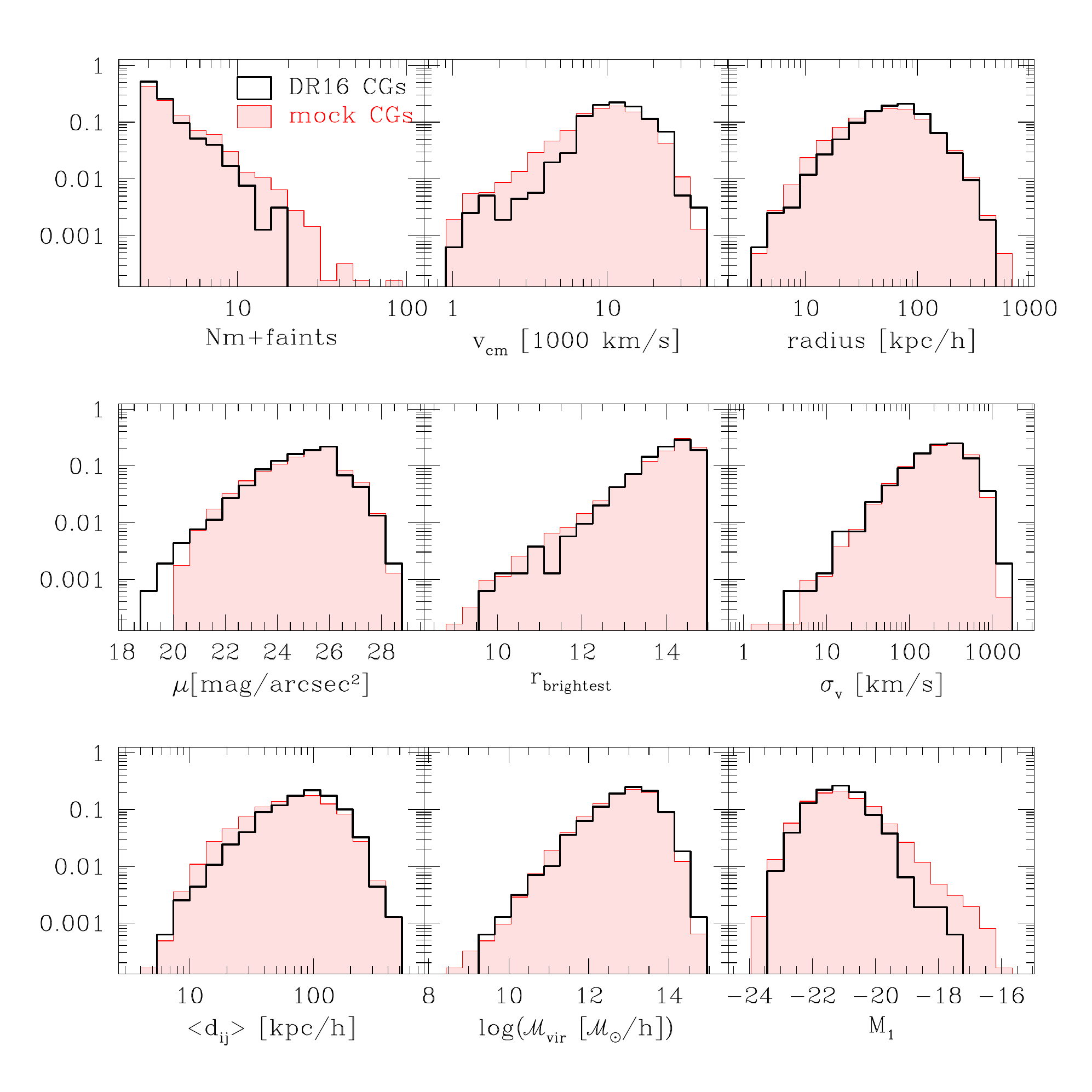}
    \caption{Distributions of properties of observational (black empty histograms) and mock (red shaded histogram) CGs. Mock CGs have been restricted to those with 3 or more observable members (after blending) and with $v_{cm}\ge 1000 \, \rm km/s$. All properties are computed using the galaxies defined as members as well as the faint galaxies in the group cylinder. From top to right to bottom the properties are: number of galaxies; median bi-weighted radial velocity of the galaxy members; projected radius of the minimum circle that encloses all the galaxies; mean surface brightness of the group in the r-sdss band; observer frame apparent magnitude of the brightest galaxy; group line-of-sight velocity dispersion; bi-weighted median of the projected separation between galaxies, group virial mass; and rest-frame ($z=0$) absolute magnitude of the brightest galaxy. }
    \label{fig:mock_cg}
\end{figure}

CGs are identified using the same algorithm applied to galaxies in the SDSS DR16 in Sect.\ref{sec:cgs}. We find 8765 mock CGs.
Given that galaxies in the lightcone are size-less, we apply a blending procedure to mimic what happens in observations. If two galaxies are separated by less than their combined half-light radii, they will be counted as one galaxy by the observers. For mock galaxies, we assigned a half-light radius based on its stellar mass by using the analytical relation provided by \cite{Lange+15} for galaxies that are bulge or disk dominated\footnote{We use a threshold of $0.7$ for the ratio between the bulge stellar mass and the total stellar mass of galaxies to define bulge or disk dominated galaxies in the lightcone.}. Then, if the angular distance between two members of the group is less than the sum of their half-light radii, we considered the members as blended and, therefore, they are counted as one single galaxy. This procedure only affects the number of members in groups. As a result, when a CG has less than three observable (after blending) members, the group is discarded from the sample. 
A sample of 6247 CGs with three or more observable members remains.
From the comparison of the samples of mock and observational CGs, we realised that a small fraction of mock CGs are found at the very close Universe, which is not seen in observations (mainly because of the small volume in observations). Therefore, we apply a radial velocity threshold of $1000 \rm km/s$ for mock CGs. 
The final sample comprises 6219 CGs. As we did in observations, we also look for faint galaxies inhabiting the isolation cylinder around each CG, i.e., within three times the angular radius of the minimum circle that encloses the galaxy members and with radial velocities within 1000 km/s from the group centre.
Figure~\ref{fig:mock_cg} shows the distributions of properties of observational (empty histograms) and mock (shadow histogram) CGs. All properties are computed using the complete sample of galaxies around CG centres (main members and faint members).

\subsection{Loose groups}

\begin{figure}
    \centering
    \includegraphics[width=8cm]{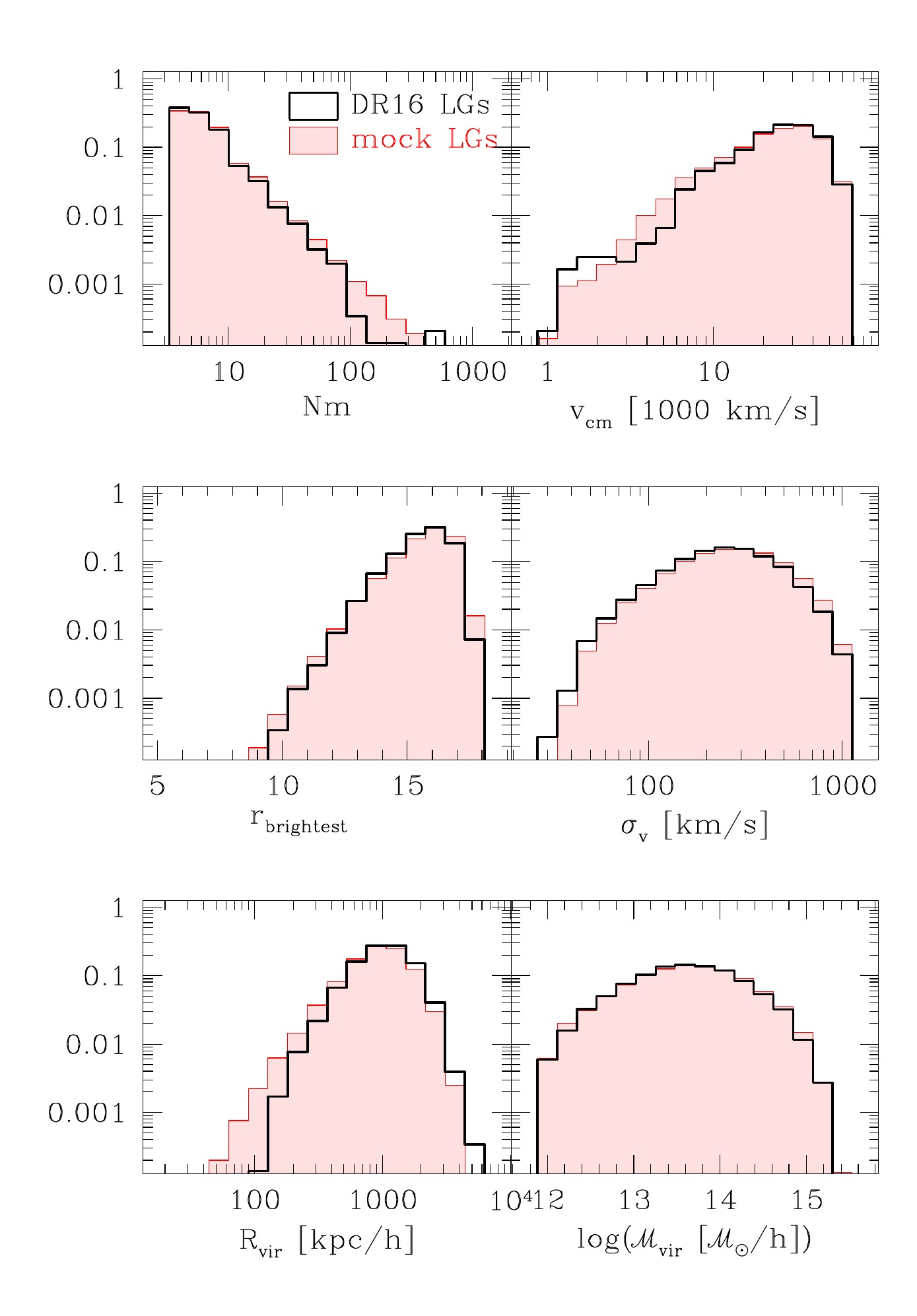}
    \caption{Distributions of properties of observational (black empty histograms) and mock (red shaded histogram) loose groups. Mock loose groups have been restricted to those with 4 or more observable members (after blending) and radial velocity greater than 1000 km/s. From top to right to bottom the properties are: number of galaxy members; median bi-weighted radial velocity of the group members; observer frame apparent magnitude of the brightest galaxy; group line-of-sight velocity dispersion; 3D virial radius; and group virial mass. 
    }
    \label{fig:mock_loose}
\end{figure}

The identification of loose groups in the lightcone was performed using the same algorithm applied to observations (Sect.~\ref{sec:grp}). We applied the blending procedure described above to galaxies in loose groups, and discarded from the sample those loose groups with less than four observable members. 
The final sample comprises 75110 mock loose groups with four or more observable members, virial masses larger than $10^{12} \ {\rm {\cal M}_{\odot} \ h^{-1}}$, and radial velocity larger than $1000 \, \rm km/s$. The latter restriction is imposed to mimic observations.
The distributions of properties of loose groups are shown in Fig.~\ref{fig:mock_loose}. Black empty histograms correspond to loose groups identified in the DR16, while red shaded histograms correspond to loose groups in the mock lightcone. 
Among these, a sample of 11893 loose groups have a first-ranked galaxy brighter than $14.77$ magnitudes and at least 3 members within a 3 magnitude range from the brightest galaxy to mimic the magnitude concordance criterion applied in CGs.

\label{lastpage}

\end{document}